\documentclass[twocolumn]{aastex62}

\usepackage{amsmath,amstext}
\usepackage[T1]{fontenc}
\usepackage{apjfonts}
\usepackage[figure,figure*]{hypcap}
\usepackage{ulem}
\usepackage{lineno}

\shorttitle{EVE-{\rhessi} DEM plus Non-Thermal}
\shortauthors{J.~M.~McTiernan et al.}

\newcommand{\goes}{\textit{GOES}}
\newcommand{\sdo}{\textit{SDO}}
\newcommand{\rhessi}{\textit{RHESSI}}
\newcommand{\eve}{{EVE}}
\newcommand{\etotal}{$\mathcal{E}_t$}
\newcommand{\ediff}{$\mathcal{E}_{diff}$}

\begin{document}

\title{The Multi-Instrument ({\eve}-{\rhessi}) DEM for Solar Flares, and Implications for Non-Thermal Emission\\}

\author{James M. McTiernan}
\affiliation{Space Sciences Laboratory, University of California,
   Berkeley, CA 94720}
\correspondingauthor{James M. McTiernan}
\email{jimm@ssl.berkeley.edu}

\author[0000-0001-8702-8273]{Amir Caspi}
\affiliation{Southwest Research Institute, Boulder, CO 80302}

\author[0000-0001-6102-6851]{Harry P. Warren}
\affiliation{Space Science Division, Naval Research Laboratory, Washington, DC 20375}

\begin{abstract}

Solar flare X-ray spectra are typically dominated by thermal bremsstrahlung emission in the soft X-ray 
($\lesssim$10~keV) energy range; for hard X-ray energies ($\gtrsim$30~keV), emission is typically 
non-thermal from  beams of electrons. The low-energy extent of non-thermal emission has only been 
loosely quantified. It has been difficult to obtain a lower limit for a possible non-thermal cutoff 
energy due to the significantly dominant thermal emission.

Here we use solar flare data from the {EUV Variability Experiment} ({\eve}) on-board the \textit{Solar 
Dynamics Observatory} ({\sdo}) and X-ray data from the \textit{Reuven Ramaty High Energy 
Spectroscopic Imager} ({\rhessi}) to calculate the Differential Emission Measure (DEM). This 
improvement over the isothermal approximation and any single-instrument DEM helps to resolve 
ambiguities in the range where thermal and  non-thermal emission overlap, and to provide constraints 
on the low-energy cutoff.

In the model, thermal emission is from a DEM that is parametrized as multiple gaussians in $Log(T)$. 
Non-thermal emission results from a photon spectrum obtained using a thick-target emission model. 
Spectra for both instruments are fit simultaneously in a self-consistent manner.

Our results have been obtained using a sample of 52 large ({\goes} X- and M-class) solar flares 
observed between February 2011 and February 2013. It turns out that it is often possible to determine 
low-energy cutoffs early (in the first two minutes) during large flares. Cutoff energies at these times are 
typically low, less than 10~keV, when assuming coronal abundances. With photospheric abundances, cutoff energies are typically $\sim$10~keV higher, in the $\sim$17--25~keV range. 
\end{abstract}


\section{Introduction}
\label{sec:intro}

Solar flare X-ray emission is commonly characterized as either
``thermal'' or ``non-thermal'' \citep{fletcher2011}. In the standard flare model \citep[e.g.,]
[]{shibata1996} the thermal component is ``soft'' X-ray emission due to 
bremsstrahlung radiation from a heated plasma, and the non-thermal component is
``hard'' X-ray emission due to bremsstrahlung from a beam of particles
(usually assumed to be electrons) accelerated in a reconnection-related process 
in the solar corona. (See, for example, the review by \citealt{benz2017}).

\begin{figure}[tbh]
\epsscale{1.00}
\plotone{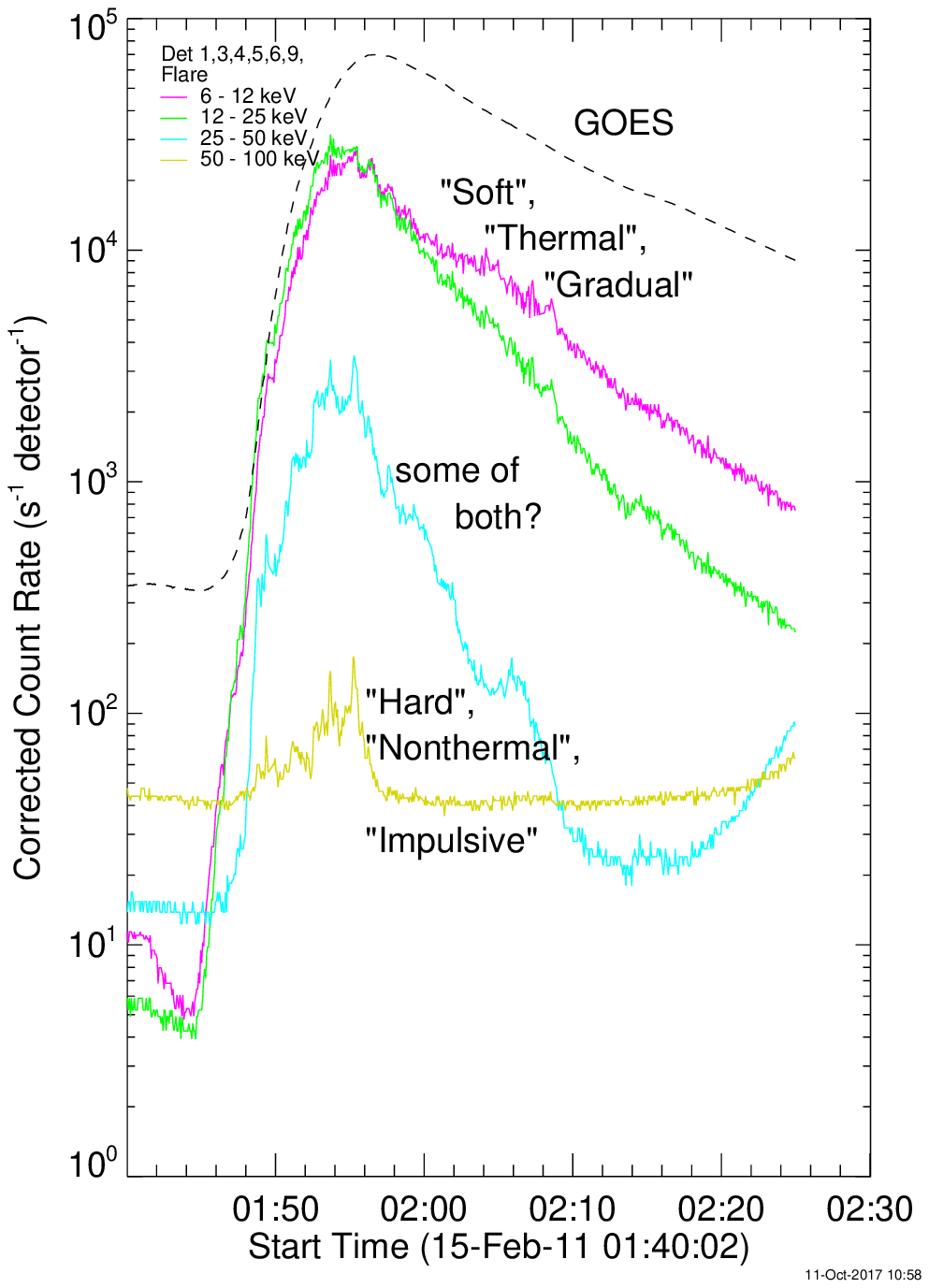}
\caption{{\rhessi} and {\goes} light curves for an X-class flare on 2011
  February 15.  {\rhessi} curves (colors) are for the energy ranges of 6--12, 
  12--25, 25--50, and 50--100~keV. The {\goes} curve (dashed) is for the long
  wavelength (nominally 1--8 {\AA}) channel.}
\label{fig:light_curve}
\end{figure}

The relative magnitudes and timing of the hard and soft components are
illustrated in Figure~\ref{fig:light_curve}, which shows X-ray
emission observed by the \textit{Reuven Ramaty High Energy Solar 
Spectroscopic Imager} \citep[{\rhessi};][]{lin2002} and by the \textit{Geostationary 
Operational Environmental Satellite} \citep[{\goes};][]{donnelly1977} X-ray Sensor from a solar flare that
occurred on 2011 February 15. The {\rhessi} light curves in the energy
bands of 6--12~keV, 12--25~keV, 25--50~keV, and 50--100~keV show
typical behavior; the lowest energy (thermal ``soft'' X-ray) emission
is gradual, the highest energy (non-thermal ``hard'' X-ray) emission
is impulsive, and the intermediate-energy emission shows both characteristics. 
Flares typically exhibit the ``Neupert Effect,'' in which the derivative of the 
gradual soft X-ray time profile is similar to the time profile of impulsive 
hard X-rays \citep{neupert1968}.

Since the thermal and non-thermal emissions overlap in energy, it 
has been a difficult problem to find a low-energy limit for the electrons
responsible for the non-thermal emission, with estimates ranging from as low as
5--6~keV \citep{kane1992} to values as high as 50~keV or more
\citep{sui2007, warmuth2009} and many values in between
\citep{aschwanden2016}. This value is important for determination of
the energy released in non-thermal electrons; a difference of just 10~keV
in this parameter can result in orders of magnitude changes in the total energy
required for the flare.

In this work, we combine {\rhessi} X-ray data with extreme ultraviolet (EUV)
data from the \textit{EUV Variability Experiment} \citep[{\eve};][]{woods2012} instrument 
on-board the \textit{Solar Dynamics Observatory} \citep[{\sdo};][]{pesnell2012} 
to calculate the Differential Emission Measure (DEM) for the thermal component 
of solar flares. This improvement over the isothermal approximation, and over 
DEMs derived from {\rhessi} alone, helps to resolve the ambiguity in the energy 
range where the thermal and non-thermal components may have similar photon 
fluxes, and hence where it is often difficult to differentiate between them 
using more na\"ive methods.

Previously we have shown that even though {\rhessi} and {\eve} are very
different instruments, they can still be used in combination to
self-consistently obtain the DEM in the $\sim$2--50~MK range \citep{caspi2014}. 
Because of the simultaneously complementary and overlapping temperature 
sensitivities of the two instruments, their joint DEM converges more stably, is 
more robust, and is significantly more accurate---particularly at the extreme 
ends of the temperature range---than when using one instrument in isolation.
In that prior work, although we did fit the non-thermal component of the 
{\rhessi} spectra, we did not consider the resulting non-thermal parameters in 
any detail.  In this work we are interested in constraining low-energy cutoffs 
in the ``residual'' non-thermal spectrum, i.e., the {\rhessi} spectrum that is 
left over after the DEM has accounted for the bulk of the soft X-ray emission. 
Ideally, we would calculate the DEM using {\eve} alone, to independently determine
the thermal X-ray emission which we would then subtract from {\rhessi} to 
obtain the residual, presumably entirely non-thermal, spectrum. However, this 
is not possible since the {\eve} DEM is not well-constrained at high 
($\gtrsim$20~MK) temperature \citep{warren2013} and the resulting predicted 
thermal X-ray component can therefore be inaccurate \citep{caspi2014}, sometimes quite significantly. 
So, as in the prior work, we will fit the DEM plus non-thermal
spectra simultaneously, with {\eve} and {\rhessi} together. We then use $\chi^2$ 
values to determine limits for the low-energy cutoff in the non-thermal 
emission.

In the following section we discuss the DEM plus non-thermal model for
the emission. This is followed by a description of the data set, then
the results of the calculation, followed by discussion of the
results.

\section{DEM plus Non-Thermal Model}
\label{sec:model}

The method we use for calculating the DEM has been presented in detail
in \citet{warren2013} and \citet{caspi2014}; here we only give a
brief overview. In the fitting model, the DEM is parametrized by a set of
11 Gaussian functions of $Log(T)$, equally spaced in the range of
$Log(T)$ from 6.2 ($\sim$1.6~MK) to 7.8 ($\sim$63~MK). The width of each 
Gaussian is $dLog(T) = 0.02$, and is held fixed in the model. Only the
Gaussian amplitudes vary. The amplitude of the 11\textsuperscript{th}
Gaussian (at $\sim$63~MK) is held fixed to a small value, to increase the
stability of the calculation, so the model DEM is effectively set to
be zero at $\gtrsim$60~MK. The original calculation \citep{caspi2014} used 10 
Gaussians and did not limit the amplitude for the highest temperature 
component; the limit on the additional Gaussian imposed here improves fitting stability and is consistent 
with prior studies that found that flare plasma temperatures do not exceed 
$\gtrsim$50~MK even for the largest recorded solar flares \citep{caspi2014a, 
warmuth2016}. Following our previous work, we use coronal abundances for the 
demonstration of the DEM calculation, but we have also done the calculations 
assuming photospheric abundances and will discuss the effects of abundance
variation, particularly for Fe.

The next step is to use the CHIANTI IDL package \citep{dere1997,landi2013} to 
calculate the thermal EUV spectral irradiance from the model DEM, for comparison with 
the {\eve} data. For {\rhessi}, the X-ray photon flux is calculated using 
\texttt{chianti\_kev}, a database of pretabulated (for speed) CHIANTI-generated 
X-ray spectra, from the IDL SolarSoft \citep[SSW;][]{freeland1998} 
\texttt{xray} package, integrated over the instrument response to recover a 
model {\rhessi} spectrum\footnote{See \url{https://hesperia.gsfc.nasa.gov/rhessidatacenter/software/installation.html}}.
Note that the original model discussed by \citet{caspi2014} separately
fit the Fe and Fe-Ni line complexes (at $\sim$6.7 and $\sim$8 keV). Here we use 
the line emission as calculated directly by the CHIANTI package.

\begin{figure}[tbh]
\epsscale{1.2}
\plotone{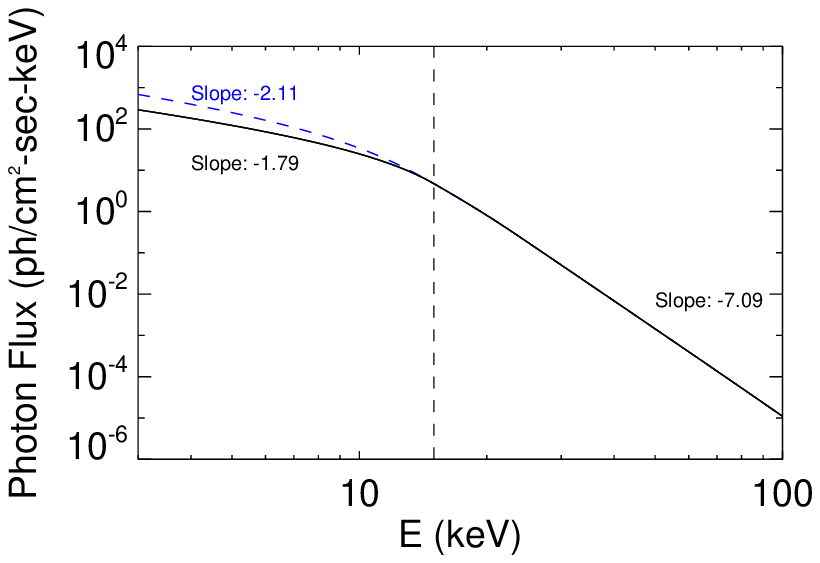}
\caption{A sample thick-target X-ray photon spectrum, for $E_c = 15$~keV 
(dashed vertical line) and input electron spectral power-law index $\delta = 
-7.8$. Note that a sharp cutoff at $E_c$ in the electron energy spectrum 
translates to a gradual rollover in the photon spectrum at energies below 
$E_c$. The blue dashed line below 10~keV shows the photon spectrum for
a model with flat electron spectrum below $E_c$, see Section~\ref{sec:discussion} for discussion.}
\label{fig:thick2_spec}
\end{figure}

To completely fit the {\rhessi} spectrum, a non-thermal emission model is
necessary in addition to the thermal (DEM) component. We use the thick-target 
model \citep{brown1971}, as implemented by \texttt{f\_thick2} in the SSW 
\texttt{xray} package. In this model, the non-thermal emission is assumed to be 
due to bremsstrahlung radiation from a beam of electrons, excited in the
corona by an undetermined process, impacting the chromosphere and depositing 
all of their energy there. Although the electrons are beamed along magnetic 
field lines, the distribution of pitch angles (defined as the angle between the 
electron velocity vector and the local magnetic field) is assumed to be 
isotropic. This is a reasonable assumption for the relatively low electron 
energies in which we are interested \citep{leach1981,mctiernan1990}. 
\texttt{f\_thick2} uses the \citet{haug1997} approximation to the relativistic 
Bethe-Heitler bremsstrahlung cross-section \citep[Eq. 3BN of ][]{koch1959}.
For an initial power-law electron beam with spectral index (negative log-log 
slope) $\delta$ and a low-energy cutoff at electron energy $E_c$, we expect to 
see a break (or, more correctly, a rollover) in the photon spectrum slightly 
below the electron cutoff energy, with the spectrum below $E_c$ being flatter 
than the spectrum above, as shown in Figure~\ref{fig:thick2_spec}. The photon
spectral index below the cutoff has an asymptotic value of $\sim$1.8,
which is independent of the value above the cutoff. In 
Figure~\ref{fig:thick2_spec} the photon index $\gamma$ above the cutoff is
$\sim$7.1 (for input electron $\delta = 7.8$).

In our model, using \texttt{f\_thick2}, the electron distribution
is parametrized as a broken power law, with both low- and high-energy
cutoffs. The non-thermal parameters are: $A_0$, the total integrated electron 
flux, in units of $10^{35}$ electrons s$^{-1}$; $A_1$, the power-law index of
the electron distribution function below a break energy $E_{br}$; $A_2$, the 
break energy $E_{br}$, in keV; $A_3$, the power-law index
above $E_{br}$; $A_4$, the low-energy cutoff $E_c$, in keV; and $A_5$, the 
high-energy cutoff, in keV. Including the 11 parameters for the DEM and these 6 
for the non-thermal model, there are 17 total model parameters.

The spectral fit procedure minimizes $\chi^2 =
\chi^2_{{\rhessi}}+\chi^2_{EVE}$, which for each instrument is defined as

\begin{equation}
\chi^2 = \sum_i (f_{i,model}-f_{i,obs})^2/\sigma_i^2 ~~~,
\label{eqn:GOF}
\end{equation}

\noindent
where $f_{i,model}$ is the model data (spectral irradiance for {\eve}, photon
count rate for {\rhessi}), $f_{i,obs}$ is the observed data, and $\sigma_i$ is
the measurement uncertainty. For {\rhessi}, the uncertainty in each
energy channel is estimated using Poisson statistics:
$\sigma_{i,RHESSI} = \sqrt{f_{i,obs}/\delta t}$, where $\delta t$ is
the time interval duration. For {\eve}, the uncertainty is given by the
observed standard deviation \citep[from calibrated Level 2 {\eve} data;][]{hock2012} of the
individual irradiance measurements during the time interval, divided
by the square root of the number of 10~s spectra averaged for the time
interval \citep{warren2013}.

For {\eve}, a pre-flare background spectrum is subtracted to isolate the flare 
emission; this is obtained for a three-minute interval immediately before 
the associated {\goes} flare start time. For {\rhessi} the process is more complicated, 
because the background level depends on spacecraft position; it increases at 
high geomagnetic latitude.  The background levels shown in 
\citet{mctiernan2009}, valid for the 2002--2006 period, have 
the appropriate latitude variation but are not applicable for the time 
intervals used here due to long-term detector changes that result
in higher overall background levels later in the mission. Here, for 
a given flare, we use the background spectrum during the nearest low-latitude 
spacecraft night interval, accounting for the long-term increase in 
background values. The spectrum is further modified for latitude variations 
using the results presented by \citet{mctiernan2009}.

\begin{figure*}[tbh]
\plotone{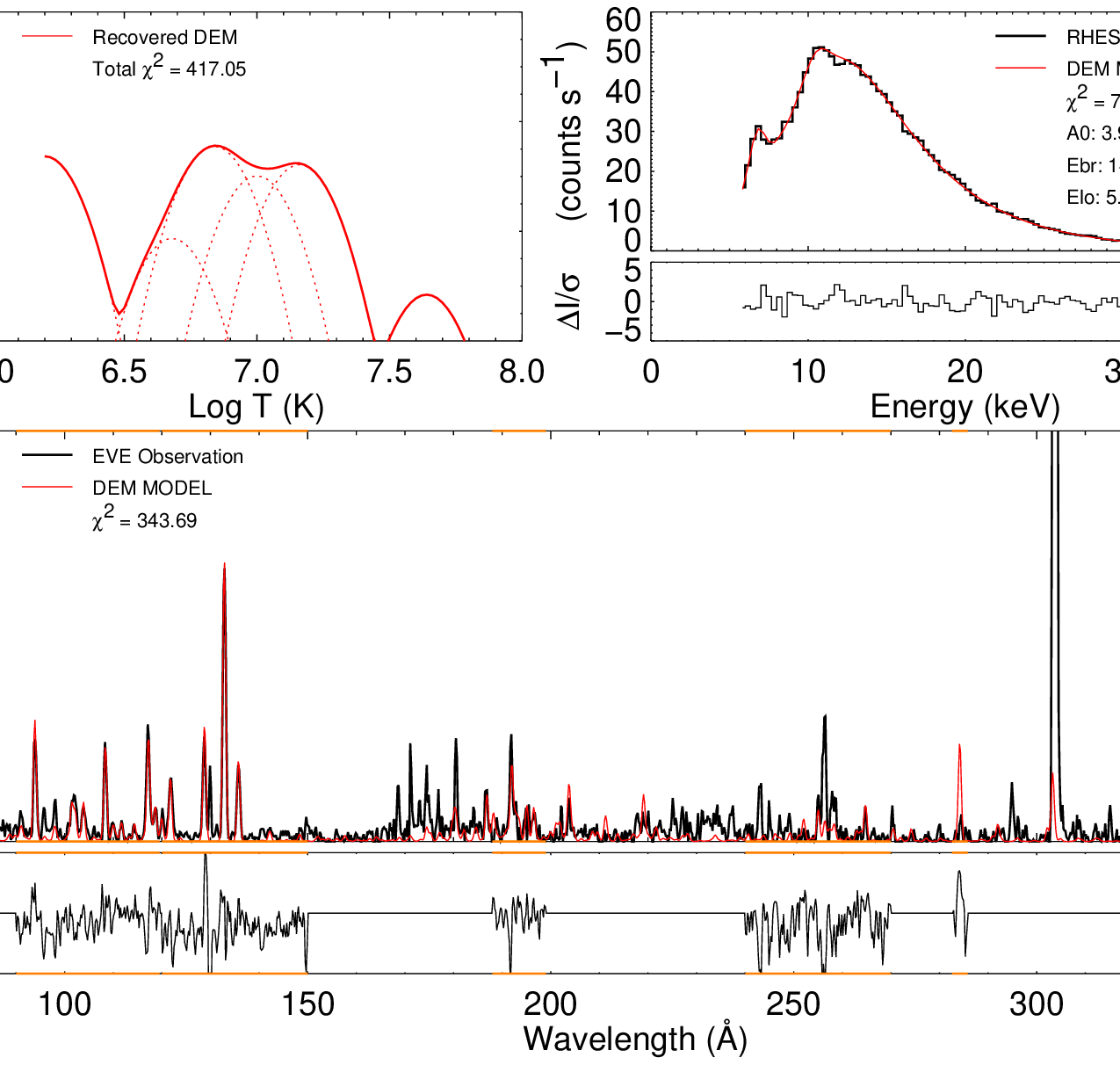}
\caption{Diagnostic plot showing the goodness of fit
  early in the M7 flare of 2011 February 13 at 17:32:36~UT. Upper
  left: Log(DEM). Upper right: {\rhessi} count rate
  spectrum and residuals. Lower: {\eve} spectrum and residuals. Orange wavelength
  ranges in the {\eve} plots include prominent flare lines and are used for fitting; 
  non-orange ranges are ignored.}
\label{fig:bigfig}
\end{figure*}

Figure~\ref{fig:bigfig} is a diagnostic plot that we use to check the
goodness of fit for the full process. The upper left corner shows the 
(recovered) model DEM. The upper right shows a comparison plot of the {\rhessi} 
count spectrum, with black denoting the observed data and red the data
expected from the model. Just below we show a plot of the residuals
for the {\rhessi} portion of the fit (normalized by the uncertainty in each 
energy channel). The lower panel shows a comparison of the {\eve} spectrum, with 
observed data in black, and model data in red. Note that not all of the {\eve}
spectrum is used for fitting; we only consider wavelength bands that include 
prominent spectral lines associated with flares (these are, mostly, relatively 
high-temperature Fe lines; see \citealt{warren2013} for more details), 
highlighted in orange. The residual values for the {\eve} spectrum are shown in 
the bottom-most panel.

\begin{figure}[tbh]
\epsscale{1.0}
\plotone{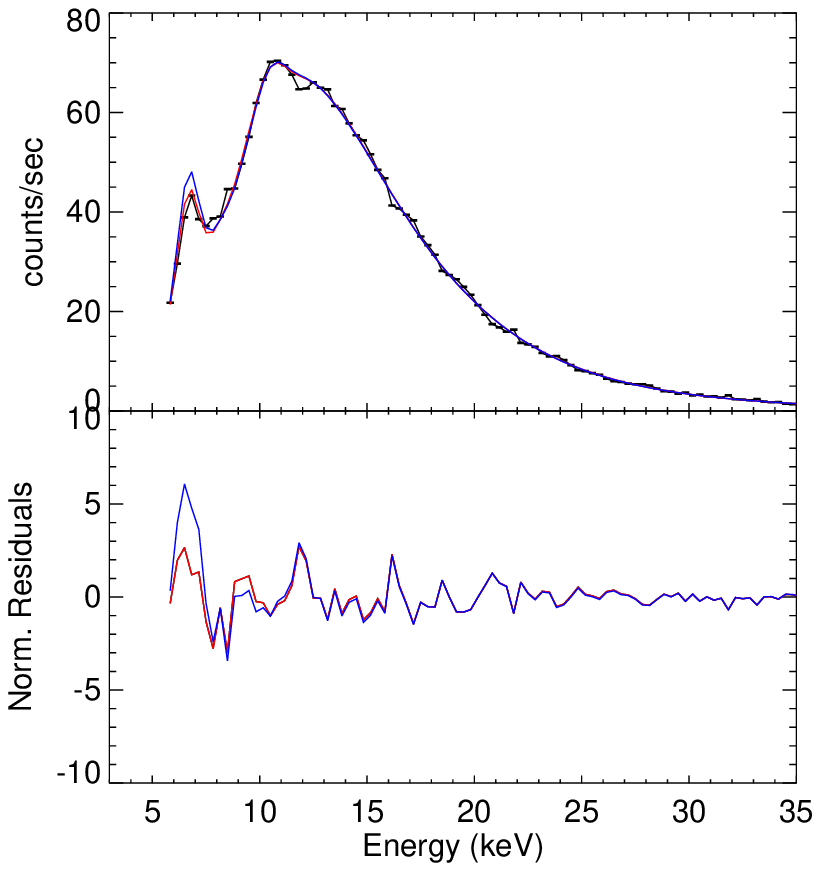}
\caption{Comparison of {\rhessi} spectral fits for the same time range
  as Figure~\ref{fig:bigfig} (2011 February 13, 17:32:36~UT).
  Top: black horizontal lines -- observed
  count rate spectrum; red line -- model results for 
  ``weighted'' fit (i.e., where {\eve} uncertainties are
  increased so {\rhessi} data more strongly influences the fit
  process); blue line -- model results for ``unweighted''
  fit. Bottom: normalized fit residuals for both cases;
  the unweighted model clearly yields a poorer fit to the {\rhessi} data at 
  lower energies (where thermal emission dominates).}
\label{fig:wt_vs_unwt}
\end{figure}

There are many more data points used in each {\eve} spectrum (400) than there
are in the typical {\rhessi} spectrum (100), so we change the
weighting of the {\eve} data in the joint fits by scaling the {\eve} uncertainties by a constant 
factor. The actual weight is calculated by finding the minimum value of 
$\sigma_{i,EVE}/f_{i,EVE}$ and re-scaling so that this value is equal to 0.04, that is, 
the uncertainty is always at least 4\% of the observed data value. This is 
done to ensure that {\rhessi} data points are sufficiently considered during the fit process;
Figure~\ref{fig:wt_vs_unwt} shows a comparison of spectra for the
limited (weighted) versus unlimited fit. As is shown in the bottom panel, 
without these limits to the {\eve} uncertainties, the {\rhessi} spectrum is 
not fit well in the full process, particularly in the 6--7~keV range. Typically, 
the scaling factor is between 1 and 5; for the example shown here the factor is 2.42.

More information regarding the inter-calibration of EVE and {\rhessi} and comparisons 
with {\goes} XRS data, not shown in \citep{caspi2014}, can be found in the appendix.

\section{Data set}
\label{sec:data}

From Figure~\ref{fig:light_curve}, we can see that the relative amount
of thermal to non-thermal emission increases over time during an event. When 
the non-thermal component is only a small a fraction of the total emission, it 
becomes difficult to constrain the low-energy cutoff  $E_c$ through spectral 
fitting. Thus, we would like to perform this calculation as early during a 
flare as possible.

\begin{figure}[tbh]
\epsscale{1.0}
\plotone{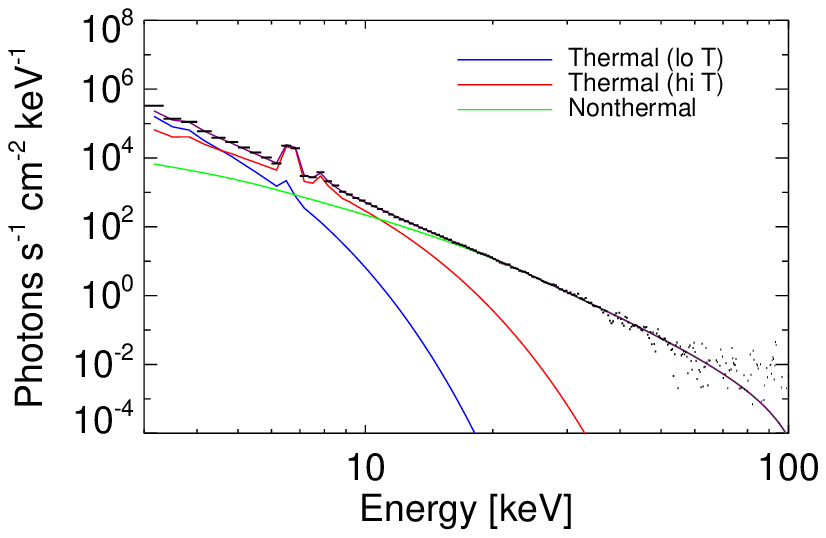}
\caption{{\rhessi} photon spectrum for 2011 February 13, 17:32:36~UT, fit 
  with two isothermal components (red, blue)
  and a non-thermal component (green). Early in the flare, the ratio of
  non-thermal to thermal emission is high, making it easier to
  estimate the low-energy cutoff of the non-thermal electron
  distribution. }
\label{fig:early_spec}
\end{figure}

\begin{figure}[tbh]
\epsscale{1.0}
\plotone{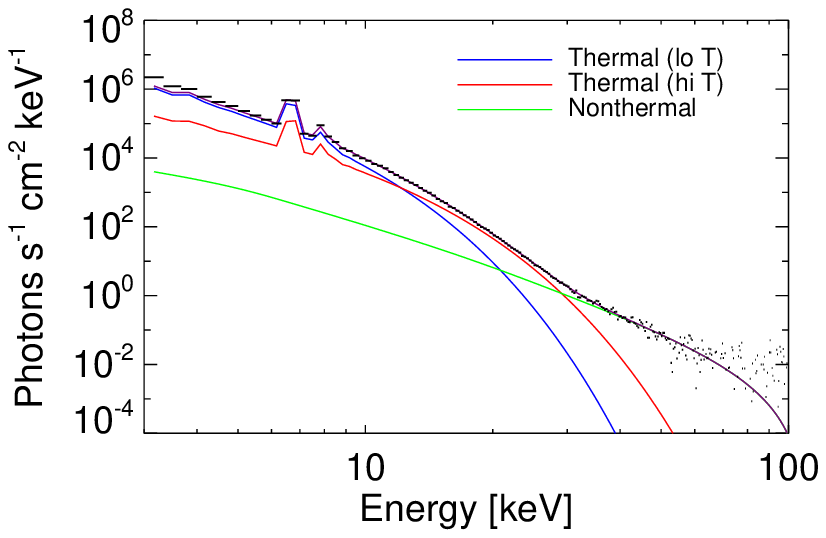}
\caption{{\rhessi} spectrum for 2011 February 13, 17:35:08~UT, again fit with 
  two isothermal components (red, blue)
  and a non-thermal component (green). At later times in an event, the
  thermal emission dominates over the non-thermal at lower energies, making the
  low-energy cutoff difficult to estimate except as an upper limit. }
\label{fig:late_spec}
\end{figure}

Figure~\ref{fig:early_spec} shows a {\rhessi} spectrum early
during a flare when the thermal component is not overwhelmingly
large. As a first cut, for demonstrative purpose, the spectrum has been fit using two isothermal
components (red, blue) and a non-thermal component (green). Even in the 
$\lesssim$10~keV range, the photon flux from the non-thermal component is not 
much smaller than that for thermal emission; this makes this time interval
a good candidate for possibly isolating a cutoff energy. Conversely, 
Figure~\ref{fig:late_spec} shows a spectrum for a later time
interval when the thermal component is much more prominent. Here, the
``crossover'' energy, where the thermal and non-thermal emissions are
approximately the same, is $\sim$30~keV. Below this crossover energy, the
non-thermal contribution to the total model becomes vanishingly small ($<$1\% for 
energies $\lesssim$15~keV), and it is clear that a spectral rollover 
corresponding to an electron cutoff $E_c$ can ``hide'' nearly anywhere under 
the dominant thermal emission below the crossover energy. Thus, $E_c$ would 
be constrained only as a relatively high upper limit, and this kind of spectrum 
is not a good candidate for finding $E_c$. For this reason, we restrict our
analysis to the first two minutes of the flares in our sample.

For this study, we work with a sample of 52 flares observed by both EVE and
{\rhessi} during the period from February 2011 to February 2013. Each
flare is of {\goes} class M or larger and has X-ray emission observed by 
{\rhessi} at energies above 50 keV. From each flare, we isolated
one or two time intervals of approximately one minute duration with good
conditions, i.e., occurring during the first two minutes of the ``impulsive''
emission $>$25~keV, with a discernible non-thermal component, with a 
relatively flat high-energy spectrum (thick-target electron spectral index 
$\delta < 9$), and with a good signal-to-noise ratio up to at least
50~keV. From these 52 flares, we found 61 appropriate time intervals in 38 flares for which 
we found good results (reduced $\chi^2 < 3$) for the DEM calculation as described
above. To establish limits on the low cutoff energy, for each time interval we 
iteratively fit the DEM plus \texttt{thick2} model for fixed values of the cutoff 
$E_c$ ranging from 5 to 30~keV. We then examined curves of the goodness of fit 
parameter, $\chi^2$, to establish lower and upper limits on $E_c$.

\section{Results for individual time intervals}

\begin{figure}[tbh]
\epsscale{1.0}
\plotone{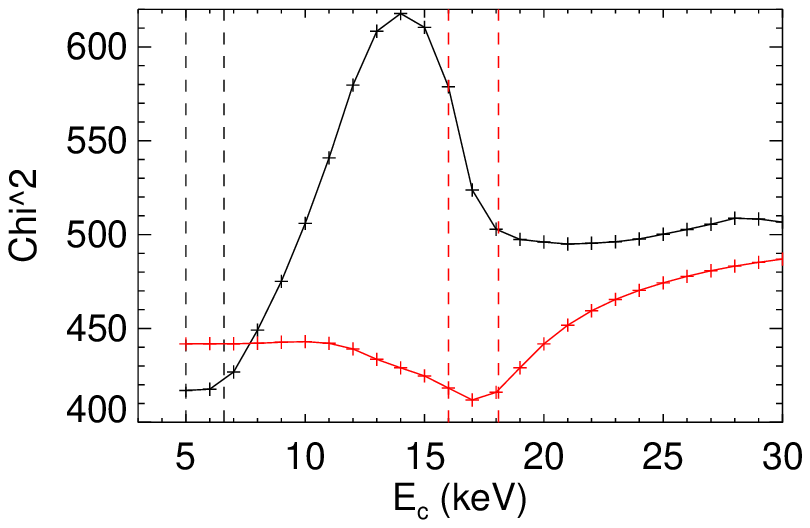}
\caption{$\chi^2$ vs. low cutoff energy $E_c$ for the time interval
  in Figure~\ref{fig:early_spec}, for coronal (black) and photospheric (red) 
  abundances. Vertical dashed lines denote upper and
  lower limits for $E_c$, defined as the points on the curve where
  $\chi^2(E_c)$ passes through $min(\chi^2)+6.63$, corresponding to the
  99\% confidence limit for the $\chi^2$ distribution.}
\label{fig:20110213_chi2}
\end{figure}

\begin{figure}[tbh]
\epsscale{1.0}
\plotone{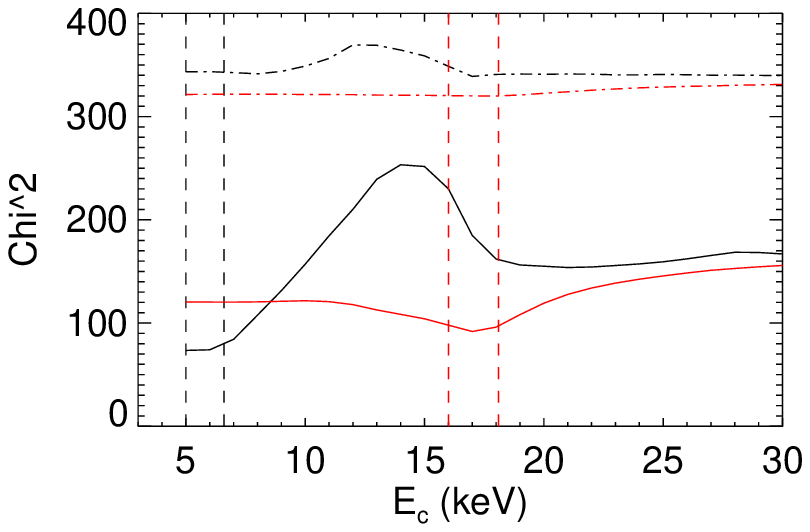}
\caption{$\chi^2_{{\rhessi}}$ and $\chi^2_{{\eve}}$ vs. low cutoff energy $E_c$ 
  for the time interval
  in Figure~\ref{fig:20110213_chi2}, for coronal (black) and photospheric (red)
  abundances. The solid lines show 
  the {\rhessi} components, while the dash-dot lines show 
  the {\eve} components.}
\label{fig:20110213_chi2_ind}
\end{figure}

Figure~\ref{fig:20110213_chi2} plots $\chi^2$ from the DEM+non-thermal 
fits as a function of the cutoff energy $E_c$, for the 
2011 February 13 flare. For this time, early 
in the flare, we have a pretty good result, and it is easy to identify upper 
and lower limits for the cutoff energy. This $\chi^2$ curve shape was typical
for most of the intervals analyzed, where for coronal abundances the lower limit is generally 5--7~keV 
and the upper limit is usually in the 8--10~keV range, and for photospheric abundances the limits are generally in the 15--18 and 18--25~keV ranges, respectively. These limits are 
determined by identifying the $E_c$ values where the (non-reduced) $\chi^2$ is less 
than 6.63 above its minimum value, which corresponds to the 99\% confidence 
limit for the $\chi^2$ distribution \citep{press1992}. 
Figure~\ref{fig:20110213_chi2_ind} shows how the total $\chi^2$ comprises the 
individual values for {\eve} and {\rhessi}. 

As can be seen from Figure~\ref{fig:20110213_chi2_ind}, 
variation in $\chi^2_{{\rhessi}}$  is much larger than that for 
$\chi^2_{{\eve}}$ so that most of the ``badness'' for high $E_c$ is in
the {\rhessi} spectrum, as would be expected since {\rhessi} is sensitive to both the thermal \textit{and} non-thermal changes caused by a change in $E_c$. Note that the values of $\chi^2_{{\rhessi}}$
and $\chi^2_{{\eve}}$ shown in Figure~\ref{fig:20110213_chi2_ind} are 
components from the combined fit, and not from fits to the spectra of the individual
instruments, and are therefore not independent. In this case the "badness" 
of the {\rhessi} spectral fit for some values of $E_c$ affects the 
fit to the {\eve} data, again expected from the joint nature of the fit.

\begin{figure}[tbh]
\plotone{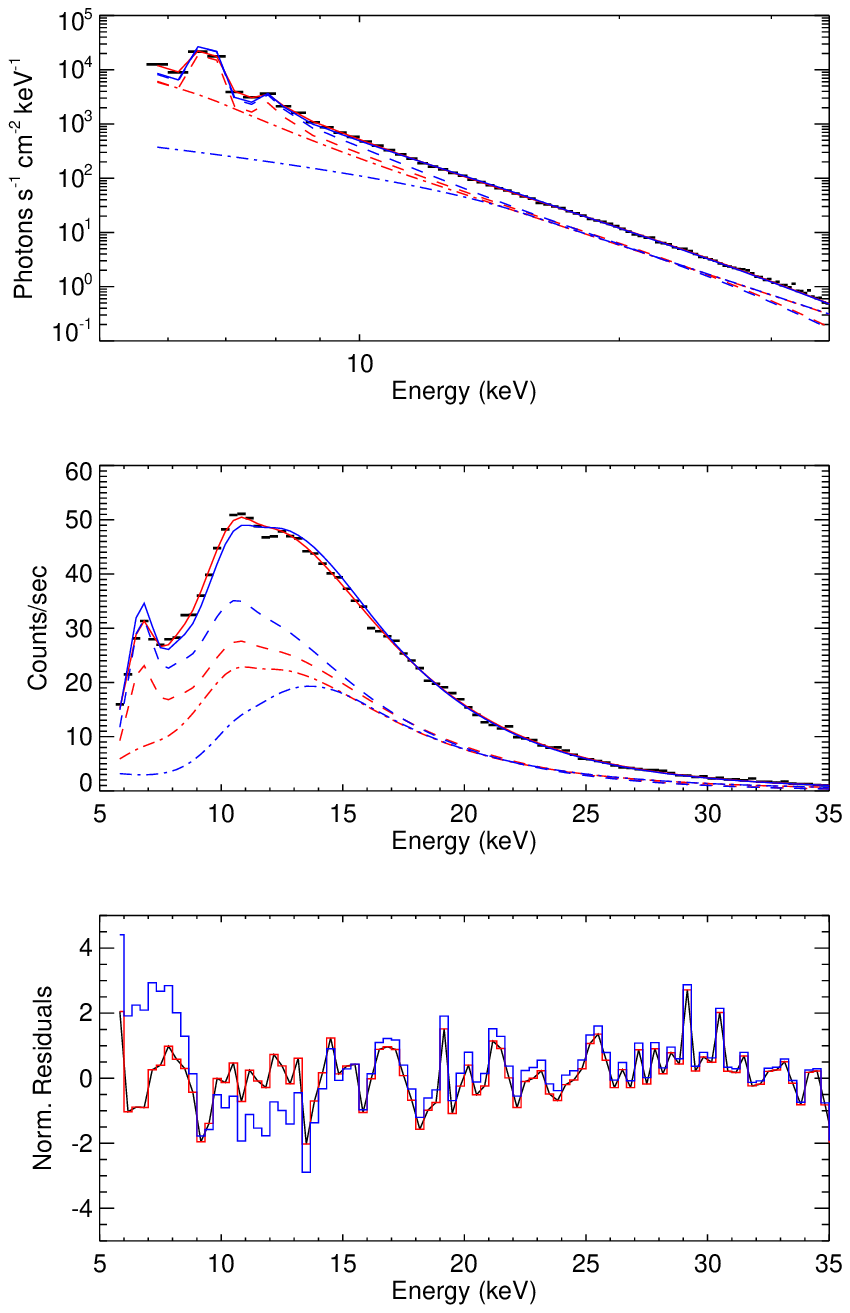}
\caption{Top: {\rhessi} photon spectra (black) for 2011 February 13, 17:32:36~UT,
with two model fits: best-fit model with $E_c$ of 7~keV (red), and a poorly-fit model with $E_c$ of 16~keV 
(blue). Dash and dash-dot lines for each color denote the thermal and
non-thermal contributions, respectively.
  Middle: count-rate spectra corresponding to above.
  Bottom: normalized residuals for the two spectral models.}
  
\label{fig:badfit16}
\end{figure}

It is instructive to examine the residuals of the spectral fits to better 
understand why a ``bad'' fit is bad, beyond just looking at $\chi^2$
values. Figure~\ref{fig:badfit16} compares the {\rhessi} count spectra
for the best fit $E_c$ value (assuming coronal abundances) of 7~keV (red) with a poorly-fit value of
16~keV (blue). For the too-high cutoff, the model spectrum does not fit the
data well in the energy range of 10--20~keV. Since the high cutoff value 
restricts how much non-thermal emission can be included in this range, 
the fit procedure tries to replace this with thermal emission. This, in turn
results in too many counts in the $\sim$6.7~keV Fe line complex which is very 
sensitive to high temperatures. In this manner, the amount of emission seen by 
{\rhessi} in the Fe line limits the allowable amount of high-$T$ emission 
measure. The requirement of fitting the {\eve} spectrum determines the amount 
of low-$T$ emission measure. Thus, there must be a substantial amount of non-
thermal emission in the 10--20~keV range, and this brackets the allowable 
$E_c$ values. The strong influence of the Fe line complex also explains the sensitivity of the best-fit $E_c$ value on the assumed abundances: photospheric abundances yield lower line fluxes for a given temperature compared to (higher) coronal abundances, hence requiring more emission measure and/or higher temperatures and therefore more overall thermal emission, pushing best-fit $E_c$ values higher.

\begin{figure}[tbh]
\plotone{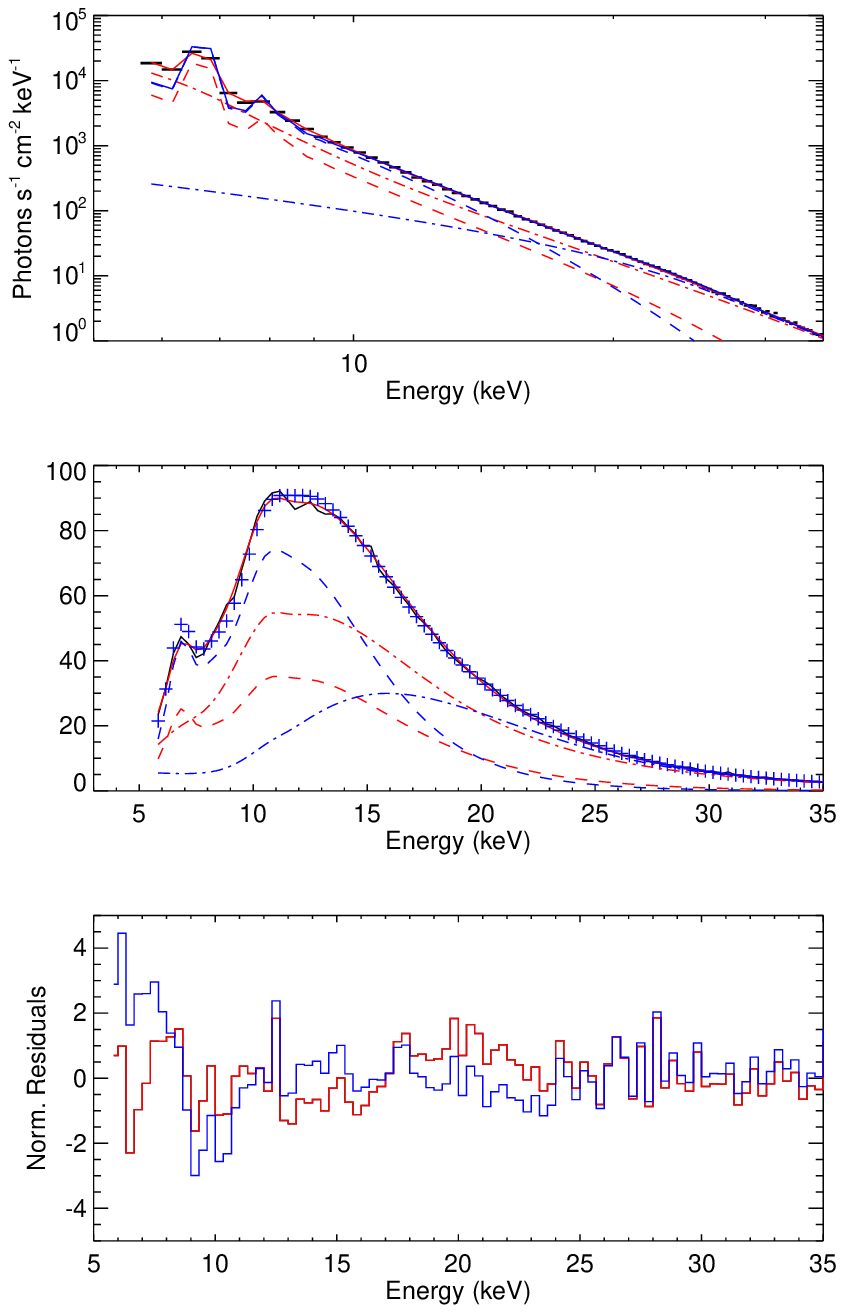}
\caption{Top: {\rhessi} photon spectra (black) for 2011 February 15, 01:48:12~UT,  
with two model fits: best-fit broken power-law model with $E_c$ of 7~keV (red), 
and best-fit single power-law model with $E_c$ of 26~keV (blue). The broken power-law model fits better in this case. Dash and 
dash-dot lines for each color denote the thermal and non-thermal contributions, 
respectively.
  Middle: count-rate spectra corresponding to above.
  Bottom: normalized residuals for the two spectral models.}
\label{fig:1pl2pl}
\end{figure}

As mentioned earlier, the spectral fits to the {\rhessi} data allow 
for a break in the non-thermal electron power law. We have also processed 
the data using a non-thermal spectrum that is a single 
power law, without a break. It turns out that the results 
are similar for the two different kinds of power-law 
spectra for most, but not all, intervals. In particular, 
approximately one-sixth (9/61) of the intervals fit the data much 
better using the broken power-law spectrum. This is 
illustrated in Figure~\ref{fig:1pl2pl}, for a time interval 
during the X flare of 15 February 2011. In this case, the 
broken power-law spectrum is a much better fit because it 
allows for an upward break in the photon spectrum. The two 
fits have similar power-law indices above 25~keV; for the 
single power law the spectral index is 5.95. For the broken 
power law, the spectral index is 5.45 above 13~keV and 
7.75 between 7 and 13~keV. As can be seen from the figure, 
the broken power law fits the data much better, 
particularly for the energy range near 10~keV. This 
upward break occurs for all of the time intervals for which the 
broken power law result differs substantially from the single power law.

Thick-target spectra with such breaks (especially upwards) are not
typically considered in flare modeling, and it is not clear what 
acceleration mechanism would result in such a spectrum. For example, it is possible
that an upward break is due to an extra component of not necessarily 
thick-target electrons with a high temperature 
$T_{electron} > T_{ion}$ which would not be considered by 
a DEM calculation based mostly on ion line emission. Distinguishing such a component
from the standard non-thermal component would
require a much more sophisticated modeling effort that is beyond the scope of this work. Here we will stay with the use of the broken  power law for the 
non-thermal spectra as this is the {\it simplest
functional form that fits the data well for most of the flare time intervals.}

Since the analysis depends on observed counts in the Fe line complex, it is
possible for results to be affected by abundance variations. Although analysis 
of {\eve} flare spectra has suggested a nearly photospheric
composition for most studied events \citep{warren2014}, analysis of
other data has yielded different results
\citep[e.g.,][]{dennis2015,doschek2017}, and this long-standing issue is still far from
resolved. In Figure~\ref{fig:20110213_chi2}, the red lines show the 
$\chi^2_{{\rhessi}}$ curve and $E_c$ limits
that result if we do the calculation using photospheric abundances, where the 
Fe abundance is approximately 4 times smaller than in the coronal case;
the $E_c$ limits derived this way are higher. 

\begin{figure}[tbh]
\plotone{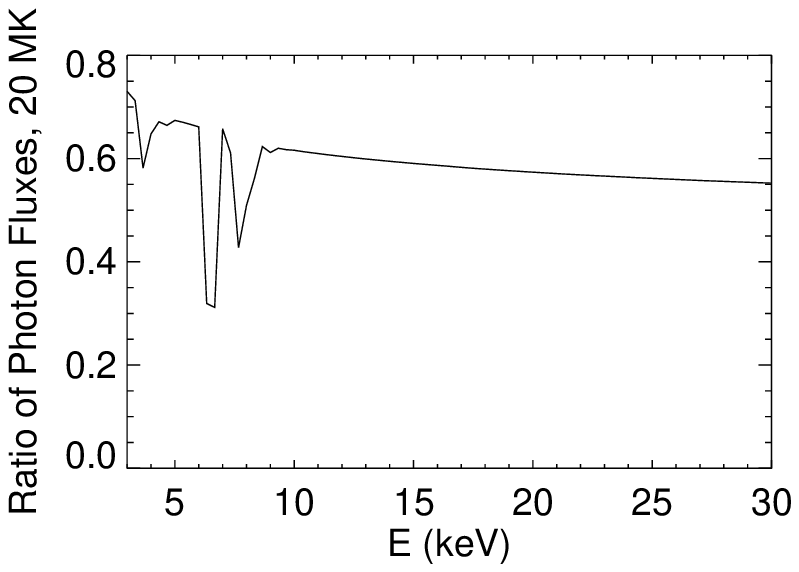}
\caption{The ratio of thermal photon fluxes for photospheric and coronal abundances for $T = 20$~MK.}
\label{fig:cvsp}
\end{figure}

The reason for this can be explained with the help of Figure~\ref{fig:cvsp}. 
This shows the ratio of thermal photon flux for photospheric abundance and
for coronal abundance for unit emission measure at 20~MK. For the Fe line 
emission in the 6--7~keV range, the ratio is approximately 0.3, reflecting the 
difference in Fe abundance. The continuum, however, is far less sensitive to abundance variations since it includes a significant contribution from hydrogen-dominated bremsstrahlung, in addition to heavier ion-dominated radiative recombination \citep{white2005}; thus, for example, at 10~keV, the ratio is only 0.6, twice as
large as for the Fe line. So, for photospheric abundance, the ratio of Fe line emission to 10-keV
continuum is approximately 1/2 the value as for coronal abundance, requiring higher model temperatures to fit the same observed line-to-continuum ratio, resulting in more thermal emission at higher temperatures and consequently higher fit values for $E_c$. The 
small values of line to continuum ratio that can be inferred from 
Figures~\ref{fig:badfit16} and \ref{fig:1pl2pl} are fully consistent 
with a substantial amount of thermal emission in the range above 10~keV, when assuming photospheric abundances, thereby
raising the cutoff energy under that assumption.

In Figure~\ref{fig:20110213_chi2} the minimum value of $\chi^2$ is 
smaller for photospheric abundance, but the difference between the two minima
($\Delta(\chi^2) = 5$) is small. For most time intervals (45 of 61), however, 
the minimum $\chi^2$ value for
coronal abundance is slightly smaller than that for photospheric abundance. (This can be seen from values for {\it reduced} $\chi^2$ are shown in table~\ref{table1}.)

In future work we will include 
elemental abundances as fit parameters which can vary during processing,
with the object of deriving rather than assuming the relative abundance values.

\section{Results for the full sample}
\label{sec:results}

Table~\ref{table1} is a comparison of the cutoff values and limits
from the {\eve}-{\rhessi} DEM models. The table columns show: interval date and 
time; derived limits for low cutoff energy $E_c$ and for the total integrated electron 
energy flux {\etotal}, using coronal and photospheric abundances; the difference between the values of {\etotal}
at the best-fit $E_c$ and with $E_c=15$~keV (for reference); values of {\it reduced}
$\chi^2$ for coronal and photospheric abundances (all of which are 
reasonably small); the ratio $R(Fe)$ of the {\rhessi} count rate in the 
$\sim$6.7~keV Fe line complex to the peak of the count rate in the 10--12~keV
range (see the middle panel of Figure~\ref{fig:badfit16}, that shows the
two peak structure in the {\rhessi} count spectrum); and the ratio $R_f(Fe)$ 
of the photon fluxes in these same energy ranges. A value of
``NA'' for $R(Fe)$ means that there was no separate peak in the count spectrum 
(because the thin attenuator was not engaged at that time), and we are unable 
to calculate that count ratio (the photon flux ratio $R_f(Fe)$ is always well defined, but
we like to use the counts ratio when possible
because it is not model dependent). 

\begin{figure}[tbh]
\plotone{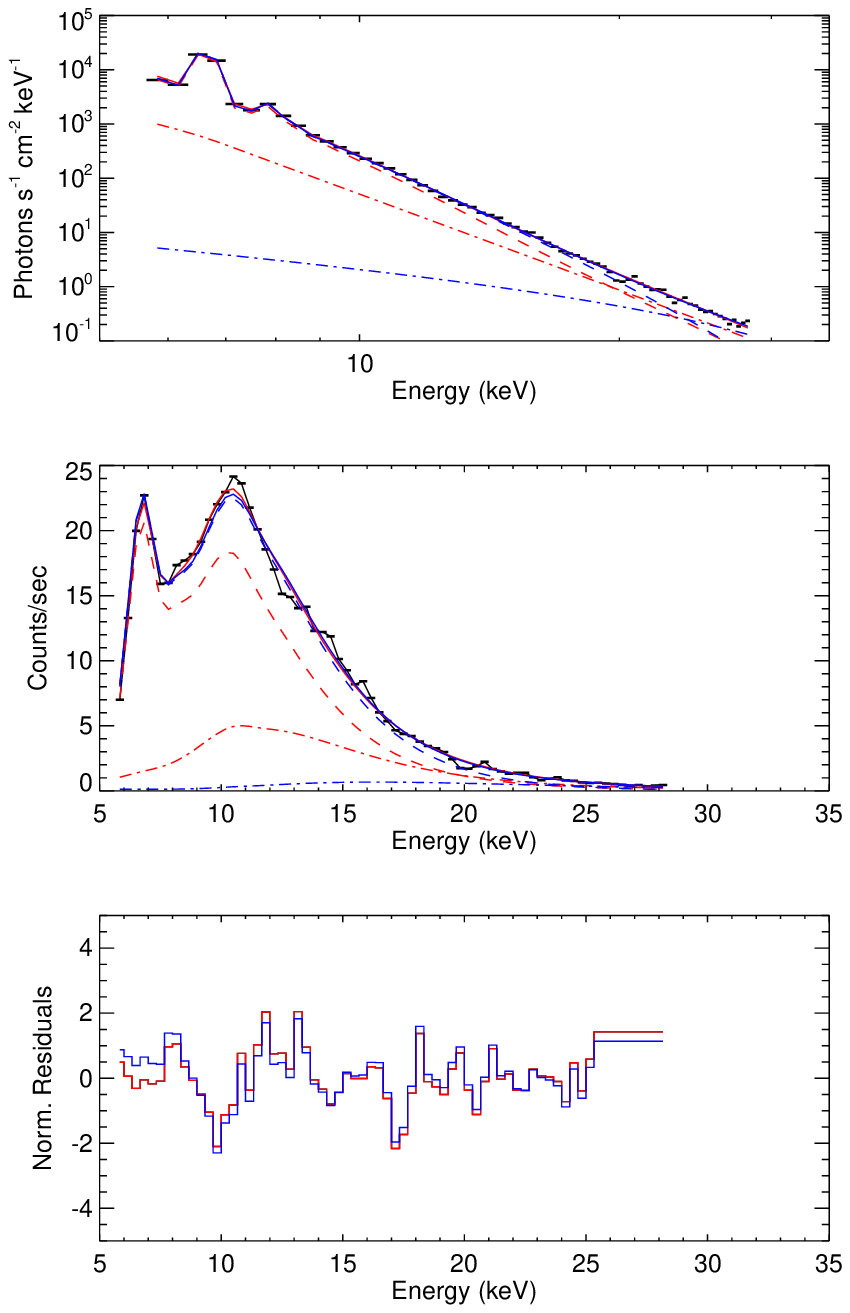}
\caption{Top: {\rhessi} photon spectra for 2011 March 9, 13:57:40~UT, 
  highlighting the effect of a relatively high Fe line-to-continuum
  ratio. Black -- observed flux; red -- best-fit model with $E_c=7$~keV; blue 
  -- model with $E_c=25$~keV.
  Dash and dash-dot lines for each color denote the thermal
  and non-thermal contributions, respectively. Middle: count-rate spectra 
  corresponding to above. Bottom: normalized residuals for
  the two spectral models. Despite the significant difference in $E_c$, the 
  two model photon spectra are nearly identical.}
\label{fig:hi_fe_ratio}
\end{figure}

For most of the examined intervals, we managed to get limiting values for 
$E_c$. For 10 of the 61 intervals, the $\chi^2$ curve using 
coronal abundances was flat or bi-modal, with a difference between high and low
$E_c$ limits greater than 15~keV, so there are 51 good sets of derived 
limits. Using photospheric abundances, there are 45 good sets of limits.

\begin{figure}[tbh]
\plotone{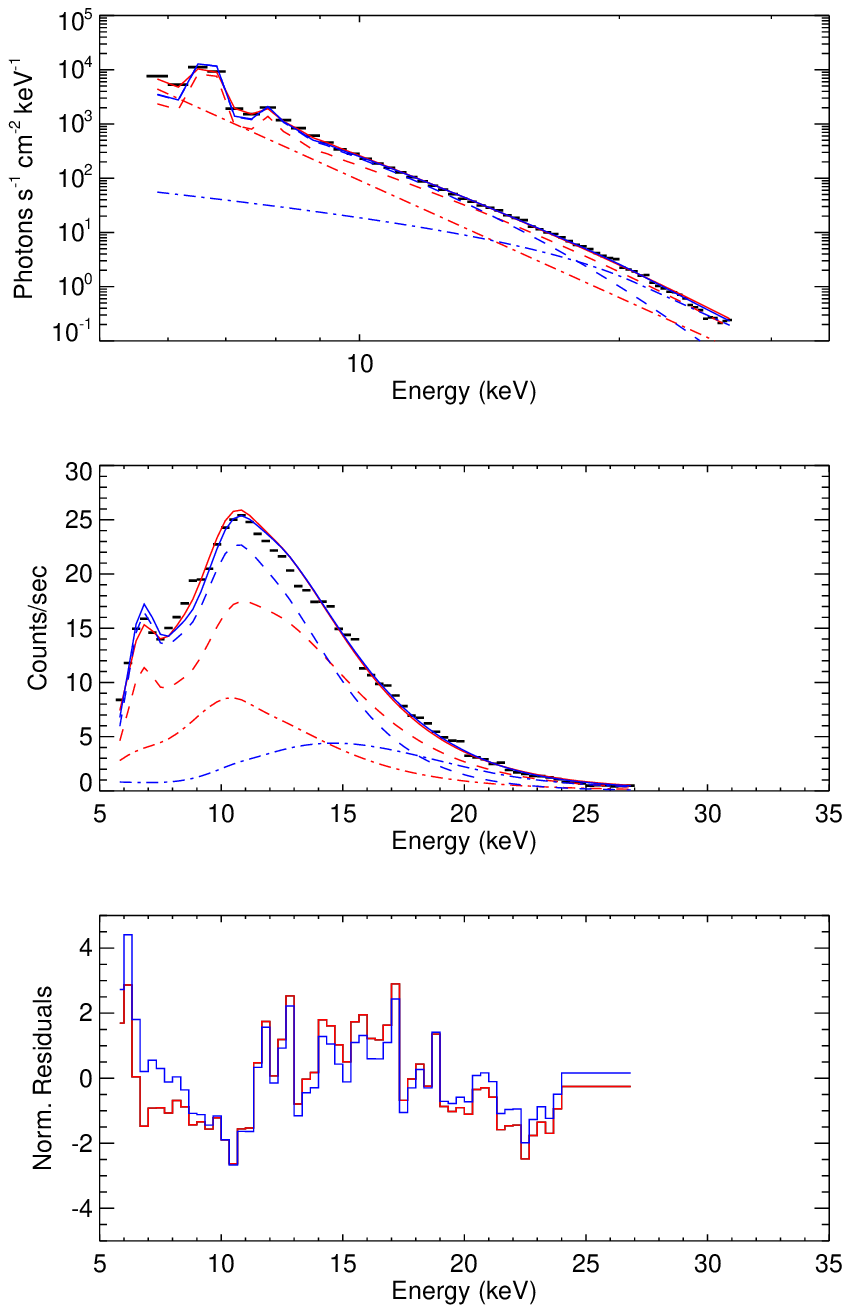}
\caption{Top: {\rhessi} photon spectra for 2011 September 6, 01:37:36~UT;  
for 
  low $E_c$ values a very steep non-thermal component 
  offsets the higher-$T$ emission measure required by a high $E_c$.
  Black -- observed flux; red -- best-fit model with $E_c=5$~keV; blue -- 
  model with $E_c=21$~keV.
  Dash and dash-dot lines for each color denote the thermal
  and non-thermal contributions, respectively. Middle: count-rate spectra 
  corresponding to above. Bottom: normalized residuals for
  the two spectral models. Despite the significant differences in the thermal 
  and non-thermal contributions for the
  two values of $E_c$, the total spectra and $\chi^2$ values for the two cases
  are similar.}
\label{fig:Bi-modal-spec}
\end{figure}

The 10 time intervals for which we obtained no good $E_c$ limits with coronal 
abundances can be divided into three different categories. (1) Six intervals
are characterized by high Fe line emission relative to the peak continuum emission
in the 10--12~keV range ($R(Fe) \gtrsim 1$), as shown in 
Figure~\ref{fig:hi_fe_ratio} and in Table~\ref{table1}. These tend to have 
very flat spectra below a break 
energy $E_{br}$ and above the cutoff value $E_{c}$, so the spectral shape does not strongly constrain $E_c$ and the resulting photon 
spectra for low and high $E_c$ values are similar. (2) Two are bi-modal in 
$\chi^2$, and for low $E_c$ values the best-fit non-thermal component is a
very steep spectrum that offsets the higher-$T$ 
emission measure required by a high $E_c$, as shown in 
Figure~\ref{fig:Bi-modal-spec}. 
(3) For the remaining two times, there is no obvious pattern in the spectra or 
fitting behavior to indicate why we cannot obtain a limit from the $\chi^2$ 
curve.
Similar patterns for photospheric abundance are not obvious.

\begin{figure}[tbh]
\plotone{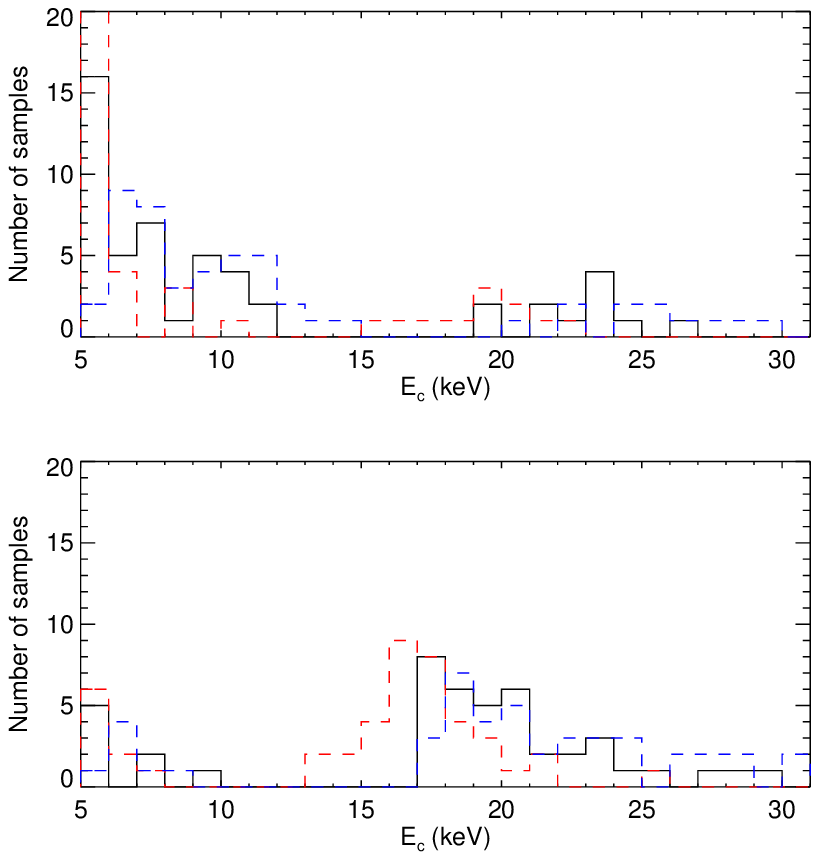}
\caption{Top: Histograms of best-fit (black), and lower (red) and upper (blue) 
limits, for the low-energy cutoff $E_c$ for the 51 time intervals with 
discernible limits using coronal abundances. Bottom: as above,
  for the 45 time intervals with good limits using photospheric abundances.}
\label{fig:histoall}
\end{figure}

Figure~\ref{fig:histoall} shows histograms of derived cutoff energies. For coronal abundance, most
of the $E_c$ lower limits (39 of 51) are less than 10~keV, and many of these intervals (32 of 51) have $E_c$ lower limits of 
5~keV, in 
the range where {\rhessi} begins to lose sensitivity \citep{smith2002}. For 
those cases, the lower limit may be less than 5~keV since we do not include 
{\rhessi} data below 5~keV in processing. In contrast, for photospheric abundance only 8 intervals have $E_c$ below 10~keV.

For coronal abundance, one-half of the intervals (26 of 51) have both upper and lower $E_c$ limits
at or below 10~keV. Only a few (12) intervals for coronal abundance have
$E_c$ lower limits above 10 keV. 
In comparison, for photospheric abundance, most intervals (30 of 45) have $E_c$ upper limits between 15 and 
25~keV, and most (36 of 45) also have lower limits for $E_c$ above 10~keV.

\section{Discussion and Conclusions}
\label{sec:discussion}

We have shown that the {\rhessi}+{\eve} DEM model generally yields values for
upper and often though less frequently) lower limits to the non-thermal low cutoff energy $E_c$, 
early in flares when the thermal emission does not overwhelm the non-thermal
component. When assuming coronal abundances, these derived cutoff energies are typically low, below 10~keV, with
most values of the lower limits in the 5--7~keV range and upper
limits below 20~keV. For
photospheric abundances, the $E_c$ values are typically $\sim$10~keV
higher.

For most of the analyzed time intervals, upper limits for $E_c$ can be
obtained because the amount of high-$T$ emission measure is strongly constrained by
the flux observed in the $\sim$6.7~keV Fe line complex. For
time intervals with a relatively large amount of Fe line emission relative to the adjacent continuum, however---roughly
1/6\textsuperscript{th} of the sample---the spectral shape is also much flatter and the Fe line is less constraining, and we do not obtain good limits for $E_c$ in those cases. Note that these "high-Fe" 
flares exhibit significant high-$T$ components at the start of the hard X-ray
emission, and thus do not fit in well with the standard flare model 
interpretation of thermal plasma being the result of ``chromospheric 
evaporation'' driven by energy deposition from non-thermal electrons. These 
may be examples of \textit{in-situ} heating as
discussed by, e.g., \citet{caspi2010}, \citet{longcope2011}, and 
\citet{caspi2015}.

\begin{figure}[tbh]
\epsscale{1.0}
\plotone{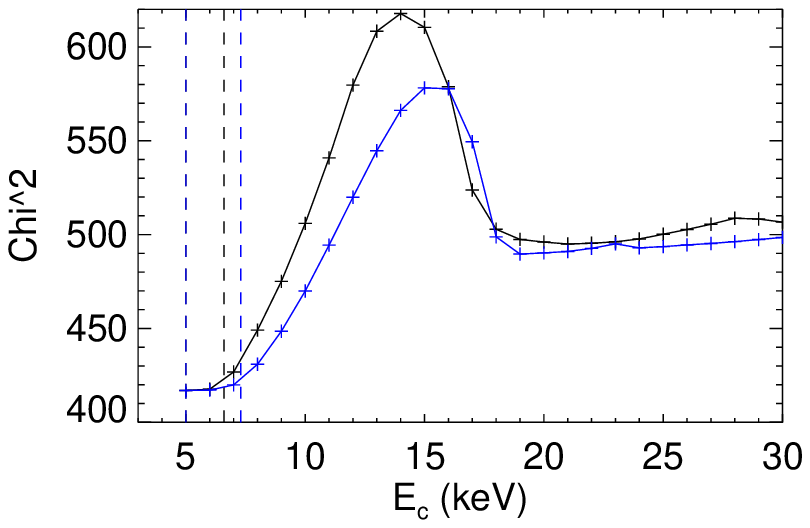}
\caption{$\chi^2_{{\rhessi}}$ vs. low cutoff energy $E_c$ for the time interval
  2011 February 13, 17:32:36~UT, for a sharp (black) and flat (blue) cutoff. 
  Dashed lines denote upper and lower limits for $E_c$. The flat cutoff model 
  has an upper limit for $E_c$ about 1~keV greater than the sharp cutoff model.}
\label{fig:20110213_chi2_flat}
\end{figure}

When we measure $E_c$, we are actually measuring \textit{the energy at which 
an assumed non-thermal photon spectrum is forced to have a downwards break
due to the presence of thermal emission.} We chose the thick-target model with a 
sharp cutoff for this work both for convenience and because it is well defined and commonly used and accepted in the community. We obtain
similar results using a more physically realistic thick-target electron model 
which has a flat electron spectrum below $E_c$ \citep[e.g.,][]{sthilaire2005}). 
We have reprocessed the sample times using a model with a flat spectrum
below $E_c$ and typically find only a 1--2~keV difference between limits obtained
using this flat cutoff model and the nominal sharp cutoff model, for most flares 
and particularly for those with low $E_c$ values. This is because the 
photon spectra for the two different models do not differ very much in the 
few keV just below $E_c$. This can be seen in Figure~\ref{fig:thick2_spec}, 
where we compare the spectra for the sharp and flat cutoff cases. Since the 
low-cutoff flares have upper limits from 7 to 10~keV, and the 
models are only fit above 5~keV, we should not expect to see much difference.
For example, in Figure~\ref{fig:20110213_chi2_flat}, we compare the $\chi^2_{{\rhessi}}$ 
curve for a sharp cutoff to that for a flat cutoff, and the difference in the 
upper limit found is 1~keV, with no difference in lower limit. For the full sample of flares, the difference between the best fit $E_c$ 
for the flat cutoff and for the sharp cutoff is less than 2~keV for 51 of 
61 intervals for coronal abundances. 

We have also tested thin-target electron models and \textit{ad hoc} 
(empirical) sharply-broken power-law photon spectra with photon spectral index 
below a given $E_c$ fixed at 1.5 (as has been commonly used in prior studies) versus the $\sim$1.8 and $\sim$2.1 gradual 
rollovers seen from the \texttt{f\_thick2} models shown in
Figure~\ref{fig:thick2_spec}. We obtain similar results for the thin-target and \textit{ad hoc} models as 
for the nominal thick-target case.

The main conclusion to take away from this work is that, for most of
the flares in the sample, it is possible to obtain limits on $E_c$
early in the flare. Values are relatively low ($<$10~keV) for coronal 
abundance because the amount of emission observed in the $\sim$6.7~keV Fe line 
complex limits the amount of high-$T$ emission measure, and therefore limits the 
thermal continuum contribution to the energy range above 10~keV. These results 
are not very dependent on the details of the non-thermal model and should be 
valid for \textit{any} model with a cutoff in the electron distribution, or that 
requires a relatively flat photon spectrum below an energy $E_c$. 

Ours are lower limits than have been determined in past calculations using 
{\rhessi} data, such as by \citet{sthilaire2005} or \citet{sui2007}, which are 
typically above 15~keV. We obtain similar $E_c$ values, above 15~keV, when using photospheric abundances.

\begin{figure}[tbh]
\epsscale{1.0}
\plotone{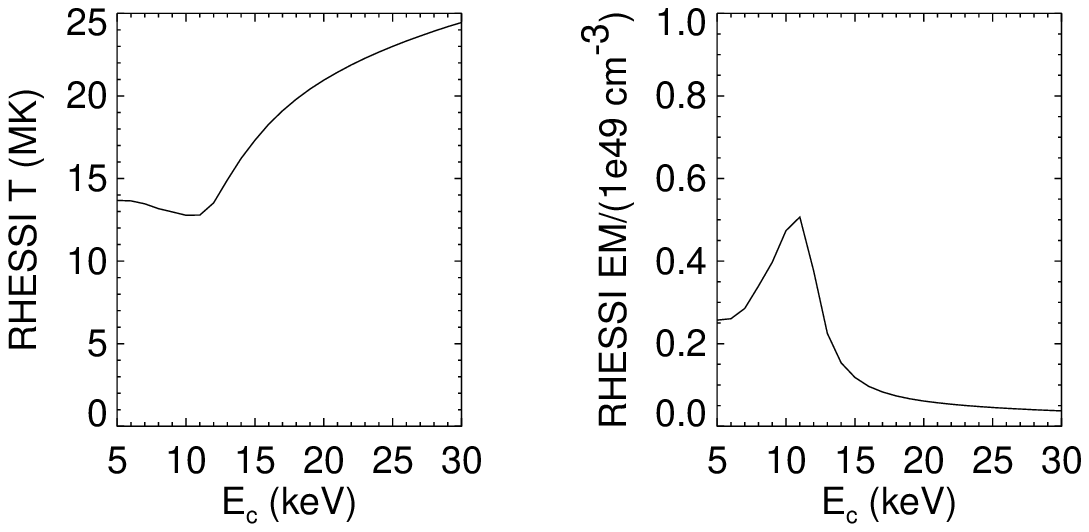}
\caption{Best-fit temperature $T$ (left, in MK) and emission measure $EM$ (right, in $10^{49}$cm$^{-3}$) values for isothermal {\rhessi} spectra as functions of low cutoff 
energy $E_c$ for 2011 February 13, 17:32:36~UT.}
\label{fig:singlet_test0}
\end{figure}

\begin{figure}[tbh]
\epsscale{1.0}
\plotone{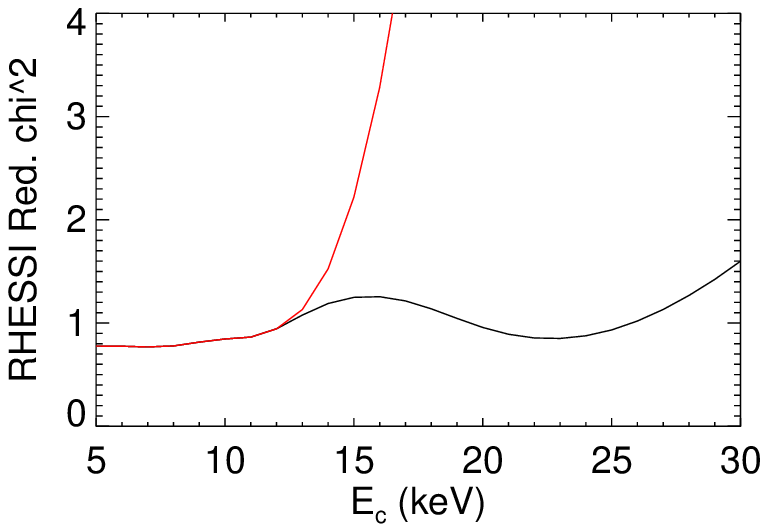}
\caption{Values of reduced $\chi^2$ for isothermal {\rhessi} spectra as 
functions of low cutoff energy $E_c$ for 2011 February 13, 17:32:36~UT. The black line shows values for spectral fits
for which the temperature was unrestricted, as in Figure~\ref{fig:singlet_test0}. The red line shows values for
spectral fits for which the temperature was capped at 14~MK.}
\label{fig:singlet_test1}
\end{figure}

We can see how these low values of $E_c$ for coronal abundance might arise
by examining just the {\rhessi} data. In a similar procedure to the {\eve}-{\rhessi}
DEM calculations, we fit RHESSI isothermal plus thick-target spectra for each
value of $E_c$ from 5 to 30~keV. Figure~\ref{fig:singlet_test0} shows the fit $T$ and 
$EM$ for these isothermal plus thick-target spectra as a function of $E_c$ for the 
time interval in Figure~\ref{fig:early_spec}. Figure~\ref{fig:singlet_test1} 
shows a plot of the value of reduced $\chi^2$ output by the 
Solarsoft \texttt{OSPEX} fitting package. From Figure~\ref{fig:singlet_test0}, we see
that as the assumed $E_c$ increases, the best-fit model $T$ also increases, up to 25~MK. As discussed previously, this is expected, since higher-$T$ emission is required by the reduction in non-thermal emission from increased $E_c$. From 
Figure~\ref{fig:singlet_test1}, we see that the $\chi^2$ curve, the black line, 
is relatively flat, so that the ``correct'' value of $E_c$ is not strongly distinguished by the $chi^2$ statistic with {\rhessi} data alone. This is unlike the $\chi^2$ curves for most of the {\eve}-{\rhessi} 
DEM models. The red curve shown in Figure~\ref{fig:singlet_test1} is for RHESSI 
spectral fits for which we have restricted the temperature to be less than 
$1.2$~keV, or 14~MK, the best-fit $T$ value for $E_c=5$~keV from Figure~\ref{fig:singlet_test0}. For those spectra, we reproduce a $\chi^2$ curve that
is similar to the {\eve}-{\rhessi} DEM case. We can conclude that we find 
low values of $E_c$ using both {\eve} and {\rhessi} \textit{because the need to 
fit both instruments simultaneously results in a DEM curve that has a much lower
average temperature (in the 10--15~MK range) than might be inferred from {\rhessi} 
alone.}

It has been pointed out that, when interpreted as due to 
cold thick-target emission, relatively low values of 
$E_c$ imply very large values for the non-thermal electron flux needed to 
account for the observed emission, i.e., the so-called ``number problem'' 
\citep{benz2017}. This remains true here for the coronal abundance cases: 
in the sixth column of Table~\ref{table1}, we show $Log($\ediff$)$, the difference 
between the amount of total electron energy flux 
{\etotal} required for the best-fit value of $E_c$, and the amount of energy flux for 
$E_c=15$~keV. For 37 of the 61 samples, this value is greater than 1.0 (i.e., we 
require more than an order of magnitude more electron energy than for the 15~keV 
cutoff).

Assuming photospheric abundances would appear to mitigate this ``number problem'', as the resultant $E_c$ values are higher, thus requiring significantly less total non-thermal energy. This assumption would seem to be supported by prior studies such as \citet{warren2014}, but other studies, such as \citet{dennis2015}, suggest that coronal abundances may be more appropriate, so there is no clear distinction between one choice versus the other. As noted previously, the higher $E_c$ values determined with photospheric abundances are also more consistent with prior studies of {\rhessi} data alone, such as \citet{sui2007}, but those studies used isothermal approximations to the thermal emission and could not consider the additional constraints on the thermal parameters afforded by requiring a simultaneous fit to a different instrument with different temperature sensitivity (in our case, EVE). We also note that fewer intervals were well fit under the photospheric abundance assumption, in part because the $\chi^2$ appears to be less sensitive to $E_c$ under this assumption, as is clear from the red curves in Figures~\ref{fig:20110213_chi2} and \ref{fig:20110213_chi2_ind}.

Other physical considerations may help to mitigate 
this ``number problem'' besides the choice of abundances, as well, such as consideration of more realistic non-thermal models including, for example, return current \citep{zharkova1995} or 
``warm-target'' plasma \citep{kontar2015,kontar2019}. Many of our lower $E_c$ limits of 5--7~keV are only a 
few times greater than the temperatures of 10--30~MK (equivalent to approximately 
1--3~keV) that we obtain in the DEM calculation, suggesting that a warm-target 
model may be most appropriate. We can make direct comparisons with results for 
some of the flares that we have analyzed with results shown by 
\citet{aschwanden2016}, albeit for different time intervals. For example, in
that work, the 2011 February 13 flare that we have been 
using for demonstration is shown to have a ``warm-target'' cutoff of 8.3~keV. 
This is close to the limits that we obtain here, using a vastly different 
calculation and more assumptions. Most of the flares analyzed in 
\citet{aschwanden2016} have low ($<$10~keV) cutoff energies for the ``warm-
target'' approximation, similar to our results here.

As noted above, the original DEM model discussed by \citet{caspi2014}
separately fit the Fe and Fe-Ni line complexes (at $\sim$6.7 and $\sim$8~keV), 
and did not use the CHIANTI package for those lines. This stemmed from 
suggestions by \citet{phillips2006} and \citet{caspi2010} that the ionization 
fraction versus temperature for the Fe line complex may not be quite correct, 
based on analysis of {\rhessi} results.  For this work, however, it is 
absolutely necessary to include the Fe line emission as modeled by the CHIANTI 
package, rather than fitting it separately, since tying the lines and 
continuum together provides the ability to constrain the high-$T$ emission 
measure and subsequent thermal continuum emission model. We use the default 
CHIANTI ionization fraction model, as implemented in CHIANTI version 7.1.3 
\citep{landi2013}.

For most of the flares in the sample, we were able to attempt the
calculation for both of the first two one-minute intervals during the start of
flare. This added an extra level of validation in those cases; if the
limits on $E_c$ are not at least similar for the two time intervals in the 
same flare, and/or there is some systematic difference between the other model 
parameters for the two intervals, then we might suspect our calculations. Note 
that each calculation is independent for each interval; we do not 
use the common practice of relating the initial conditions for a subsequent 
time interval to those from the previous time interval in a given flare. For 
coronal abundances, for 18 of 22 flares for which good fits were obtained for 
both intervals, the limits overlap. 
This 
suggests that $E_c$ may be relatively stable on the scale of minutes, but in 
the absence of good limits over the entire evolution of multiple flare 
impulsive phases, we hesitate to draw a general conclusion.

It is important to note that, since we are looking at time intervals 
early during flares, the Fe~\textsc{xxv} ions mostly responsible for the 
$\sim$6.7~keV line complex may not be in equilibrium, i.e., the ion population may not fully reflect the balance of ionization states expected from the temperature distribution. The equilibrium time 
scale is dependent on the density of the hot plasma. Simple order-of-magnitude estimates for the 
high-temperature plasma density based on the size of the {\rhessi} image and 
the high-$T$ emission measure \citep[cf.][]{caspi2010th,caspi2014}
give densities of order $10^{9-11}$~cm$^{-3}$, as shown in Table~\ref{table2}.
\citet{phillips2004} suggested that for densities of $10^{10}$~cm$^{-3}$ or 
less, the ionization equilibrium may be problematic. However, we believe that 
ionization equilibrium is mostly attained since, for most of the flares, 
the ratios of counts and photon flux for the Fe line complex to the 
10--12~keV continuum, $R(Fe)$ and $R_f(Fe)$, remain relatively stable from 
minute to minute during the flares for which we fit two intervals, as 
can be seen in Table~\ref{table2}. (Note that this sample contains time 
intervals for which we had no good $E_c$ limits). For 15 of 22 of these flares, the 
ratio $R_f(Fe)$ varies by less than 50\%. This would not be the case if the 
Fe~\textsc{xxv} ionization had to ``catch up'' over time.

Table~\ref{table2} shows the results for the flares for which we have two
measurements, for high-$T$ emission measure $EM_{16}$ (the integral of the DEM for all 
$T >16$~MK), volume $V_{16}$, density $N_{16}$, ionization time scale $\tau_{Fe
\textsc{xxv}}$, and heating time scale, $\tau_\textsc{heat}$. The ionization time scale
$\tau_{Fe\textsc{xxv}} = 1/(N \times Q)$, where $Q$ is the ionization rate for 
Fe~\textsc{xxiv} to Fe~\textsc{xxv} \citep{jordan1970,phillips2004}. The value 
of $Q$ depends on temperature; for this case, since we are integrating over 
``all'' high-$T$, we used an average of the values for 
$T = 10^{7.2-7.5}$, or $Q=10^{-11.12}$. The heating time scale is given 
by $\tau_\textsc{heat} = ((1/EM_{16}) \times \delta (EM_{16})/\delta t)^{-1}$. 

For all of the flares in the sample $\tau_{Fe\textsc{xxv}}$ is less (usually much less) 
than $\tau_\textsc{heat}$, consistent with prior studies of early-flare emission \citep[e.g.,][]{caspi2010} and again suggesting that non-equilibrium ionization is not an 
issue for these events. The plasma density $N_{16}$ would have to be an order of magnitude lower for the ionization scale $\tau_{Fe\textsc{xxv}}$ to be as long as the heating time scale $\tau_\textsc{heat}$. 

The calculation that we have done here for $E_c$ is as comprehensive
as we can get considering the available instrumentation. For future
work it would be useful to have good observations of high-$T$ line
emissions that can be separated from the need to account for the
non-thermal continuum, so that the entire thermal X-ray continuum can be 
estimated independently and subtracted, yielding the expected non-thermal component 
directly. The \textit{Yohkoh} Bragg Crystal Spectrometer, with
Ca~\textsc{xix}, Fe~\textsc{xxv}, and Fe~\textsc{xxvi} channels, is a good 
example of one such instrument useful for DEM analysis 
\citep{culhane1991,mctiernan1999}. Soft X-ray spectra from, e.g., the 
\textit{Miniature X-ray Solar Spectrometer} \citep[\textit{MinXSS};]
[]{mason2016,moore2018} CubeSat, will provide additional diagnostics of the DEM 
and abundances for a number of elements \citep[e.g.,][]{caspi2015a,woods2017}. 
Combining \textit{MinXSS} data with that from other instruments, particularly 
{\rhessi}, will open new areas of parameter space to further constrain $E_c$ 
in different ways.

\acknowledgments This work was funded by NASA Heliophysics Guest 
Investigator grants NNX12AH48G, NNX15AK26G, and 80NSSC19K0287. AC was also partially 
supported by NASA grants NNX14AH54G, NNX15AQ68G, and NNX17AH38G; JMM was partially 
supported by the {\rhessi} project, NASA contract NAS5-98033. We thank B.R.~Dennis and R.A.~Schwartz for useful discussions. All thermal 
emission is modeled using the CHIANTI software package 
\citep{dere1997,landi2013}.

\newpage

\section*{Appendix: Calibration Notes}

\begin{figure}[tbh]
\epsscale{1.0}
\plotone{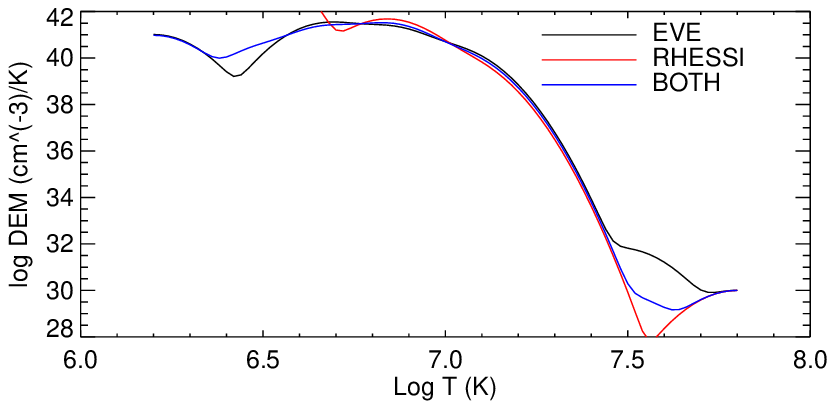}
\caption{DEM curves for {\eve} (black), {\rhessi} (red), and {\eve}+{\rhessi} (blue) for the time 
interval 2011 February 16, 14:40:44~UT to 14:41:44~UT, when no {\rhessi} attenuators were engaged.}
\label{fig:plot0_att0}
\end{figure}

For comparison purposes, it is useful to calculate results for the individual instruments.
As shown by \citet{caspi2014}, we find good agreement between {\eve} and {\rhessi}, individually, 
in the temperature range of 10--20~MK where both instrument responses overlap.  Here we compare {\eve} 
and {\rhessi} during the decay phase of some flares during February 2011. The {\eve} and 
{\rhessi} data are first fit separately (in contrast to the situation discussed regarding 
Figure~\ref{fig:20110213_chi2_ind}, where $\chi^2_{{\rhessi}}$ and $\chi^2_{{\eve}}$ 
are separate components of a combined fit). The individual DEM functions are then compared
with the DEM from the combined fit. In 
Figure~\ref{fig:plot0_att0} we show the DEM for the interval 2011 February 16,  
14:40:44~UT to 14:41:44~UT, for each instrument individually and for the combined instruments. 
The red curve, for 
{\rhessi}, and black curve, for {\eve}, are similar in the range $Log(T)$ of 6.5 to 7.4. The blue curve, 
for the combined instruments, stays with the {\eve} curve at low~$T$ and the {\rhessi} curve at 
high~$T$. 

\begin{figure}[tbh]
\epsscale{1.0}
\plotone{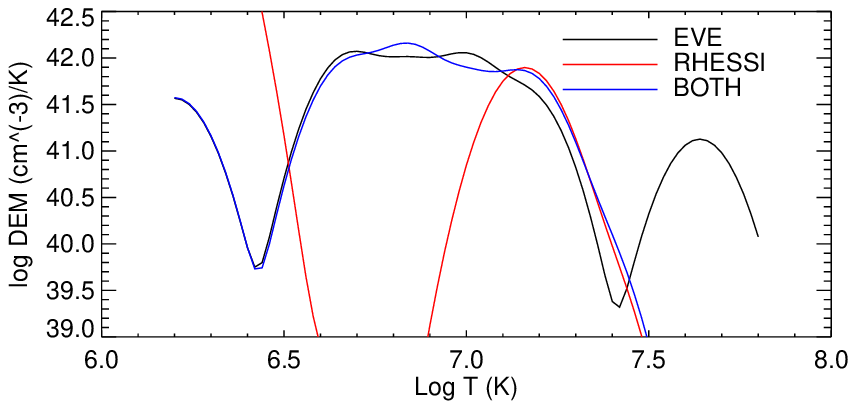}
\caption{DEM curves for {\eve} (black), {\rhessi} (red), and {\eve}+{\rhessi} (blue) for the time 
interval 2011 February 13, 17:50:32~UT to 17:51:32~UT, when the {\rhessi} thin attenuator was engaged.}
\label{fig:plot0_att1}
\end{figure}

Figure~\ref{fig:plot0_att0} shows an interval for which the {\rhessi} attenuators were out. Most of the 
intervals that we have looked at in this work, however, have the thin attenuator in. Figure~\ref{fig:plot0_att1} 
shows the DEM for an attenuator-in interval, for 2011 February 13, 17:50:32~UT to 17:51:32~UT. Here the {\eve} 
and {\rhessi} curves are similar in the range $log(T)$ of 7.0 to 7.4, with the combined solution again 
following the {\eve} curve at low~$T$ and {\rhessi} at high~$T$, with a "crossover" point, where the
individual {\eve} and {\rhessi} curves match at $log(T) = 7.1$. 

\begin{figure}[tbh]
\epsscale{1.0}
\plotone{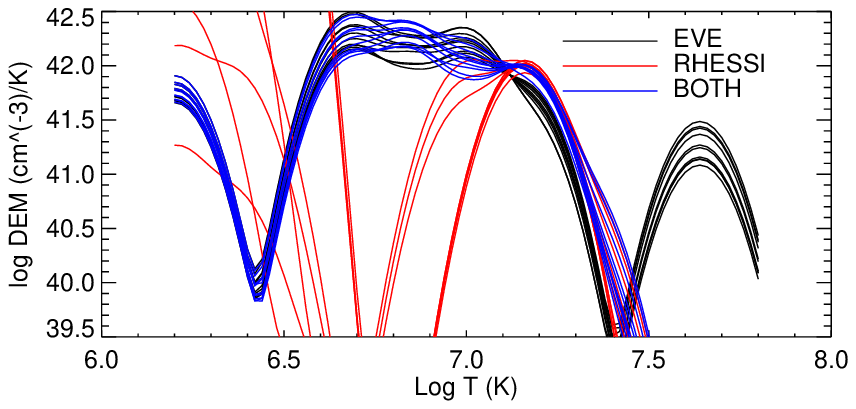}
\caption{Multiple DEM curves for {\eve} (black), {\rhessi} (red), and {\eve}+{\rhessi} (blue) for 10 different
one minute time intervals in the ranges of 2011 February 13, 17:50:32~UT to 17:55:32~UT and 2011 February 15, 02:15:48~UT
to 02:20:48~UT. For each time interval, for purposes of comparison, the three solutions are normalized so that the {\eve} DEM is the same at 
$Log(T) = 7.1$.}
\label{fig:plot2_att1}
\end{figure}

Figure~\ref{fig:plot0_att1} highlights the need to consider the combination of both {\eve} and {\rhessi} due to the poor results at either end of the temperature range when considering the instruments individually.
This is typical behavior, as shown in Figure~\ref{fig:plot2_att1}.
Here we show 10 different comparisons, 
for two sets of five 1 minute intervals during the decay phases of flares on 2011 February 13 and 15. To
remove time dependence in this comparison, for each set of {\eve}, {\rhessi} and {\eve}+{\rhessi} solutions, we normalize 
so that the {\eve} curve passes through the same point at $Log(T) = 7.1$. For all of these solutions, 
the combined (blue) curve matches well with {\eve} at low~$T$, with {\rhessi} at high~$T$, and with both in 
the middle. The high $log(T) = 7.6$ component (40~MK) that often appears in the {\eve} solutions is suppressed
when {\rhessi} is included, and the low $Log(T)<6.7$ component (5~MK), for {\rhessi}, which does not reflect 
reality, is likewise suppressed when {\eve} is included. 

\begin{figure}[tbh]
\epsscale{1.0}
\plotone{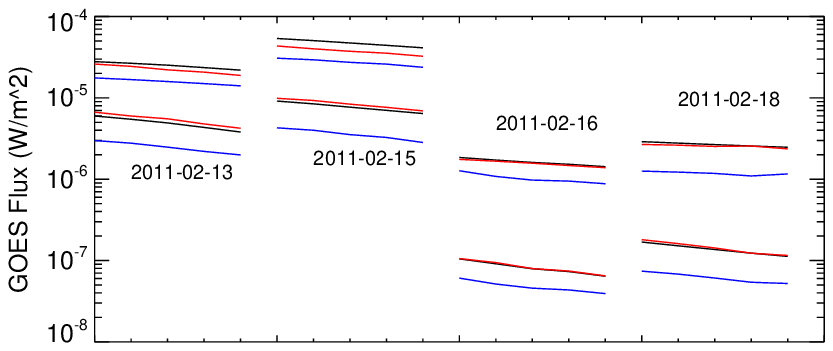}
\caption{Comparison of {\goes} flux (W~m$^{-2}$) estimates for DEM models calculated using {\eve} and {\goes} (red), and {\eve} 
and {\rhessi} (blue) during the decay phases of four flares during February 2011, compared with measured values (black). The time intervals are:
2011 February 13, 17:50:32~UT to 17:55:32~UT; 2011 February 15, 02:15:48~UT to 02:20:48~UT; 2011 February 16, 14:38:44~UT to
14:43:44~UT; and 2011 February 18, 13:15:00~UT to 13:20:00~UT.}
\label{fig:eve_goes}
\end{figure}

\citet{warren2013} found good agreement between the {\eve} and {\goes} XRS instruments. We can do a similar 
comparison here. In Figure~\ref{fig:eve_goes} we compare estimates of {\goes} fluxes (black lines) from DEM 
models including {\eve} plus {\goes}, as in \citet{warren2013} (red), and {\eve}+{\rhessi} (blue). The
{\eve}+{\rhessi} DEM underestimates the {\goes} emission by a factor of approximately 2. This may be due to
a loss of sensitivity for {\rhessi} later during the mission, as discussed by \citet{mctiernan117}.

\begin{figure}[tbh]
\epsscale{1.0}
\plotone{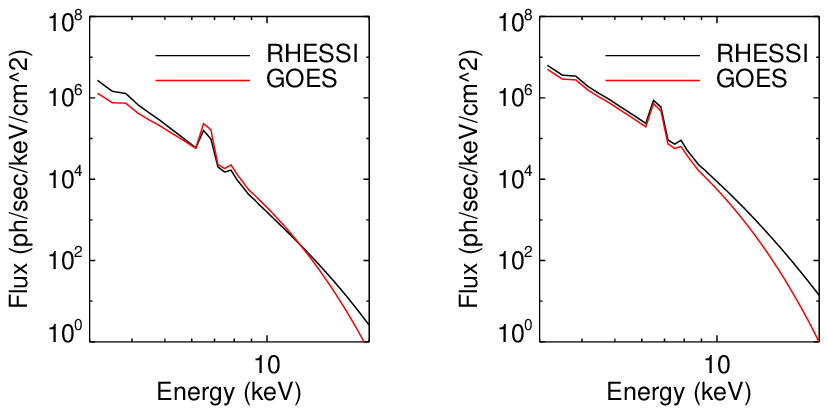}
\caption{Comparison of photon flux (photons~s$^{-1}$~cm$^{-2}$~keV$^{-1}$) for {\rhessi} spectral fits (black) 
as in \citet{caspi2010} with photon flux inferred from {\goes}-derived $T$ and $EM$ values (red) calculated via SolarSoft, for 2011 February 13, 17:50:32~UT to 17:51:32~UT (left) and 2002 July 23, 01:05:00~UT to 01:06:00~UT (right).} 
\label{fig:spex_goes_comp0}
\end{figure}

To investigate this, Figure~\ref{fig:spex_goes_comp0} shows a different sort of {\rhessi}-to-{\goes} comparison. The left-hand plot
shows the photon flux for a {\rhessi} spectral fit for an interval from the decay phase of an X-flare 
for the time interval 2011 February 13, 17:50:32~UT to 17:51:32~UT. The right-hand plot is the same comparison for 
the time interval 2002 July 23, 01:05:00~UT to 01:06:00~UT when the {\rhessi} detectors were at full sensitivity. 
In each plot the red line is the expected photon flux from a {\goes} temperature measurement for the same time 
interval. The {\rhessi} spectrum was fit using the same procedure as \citet{caspi2010}, and the {\goes} spectrum 
was calculated using the SolarSoft IDL software package. As can be seen, relative to the {\goes} flux, the 
{\rhessi} flux is less for the later time interval. We have done this comparison for all of the four time 
intervals shown in Figure~\ref{fig:eve_goes}. For the first two (with attenuator state in), for 2011
February 13 and 15, we compare with five one minute time intervals from 2002 July 23, 01:05:00~UT to 01:10:00~UT. 
The 2002 July 23 intervals were chosen to have similar {\rhessi} and {\goes} temperature values to the 2011
February 13 and 15 intervals. For non-attenuated 2011 February 16 and 18 times, we compare with 2002 July
26, 00:37:00~UT to 00:42:00~UT. 
For each interval, we calculated the ratio of {\rhessi} flux to {\goes}-derived flux in the range between 6 and 10 keV.
For the 2002 July 23 time interval, we found the ratio to be 1.36, i.e., the {\rhessi} flux is a factor of 1.36
greater than the flux calculated from the {\goes} temperature and emission measure; for the 2011 February 13 and 15 time 
intervals, the ratio is 0.78. Thus, in 2011 {\rhessi} seems approximately 0.57 as sensitive as in 2002.
The change in ratio is similar for the non-attenuated case; for 2002 July 26 the ratio is 3.11, while 
for the 2011 February 16 and 18 flares the ratio is 1.77, again showing a relative sensitivity of 0.57. Note 
also that extending the calculations done by \citet{mctiernan117} for the full mission estimates a loss of sensitivity of approximately 0.70.

\begin{figure}[tbh]
\epsscale{1.0}
\plotone{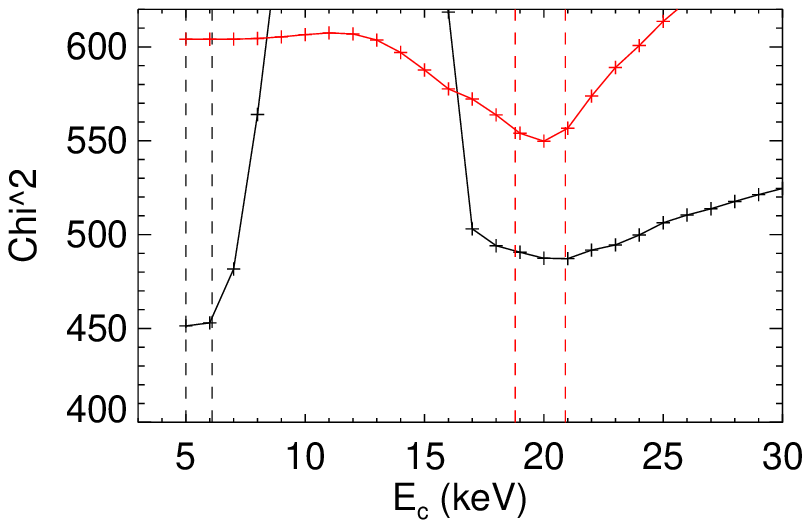}
\caption{$\chi^2$ vs. low cutoff energy $E_c$ for the time interval
  in Figure~\ref{fig:early_spec}, for the reduced-sensitivity
  models. Dashed lines denote upper and lower limits for $E_c$. For coronal abundance (black) the reduced-sensitivity
  model has an upper limit for $E_c$ about 0.5~keV less than full sensitivity model. For photospheric abundance, the reduced sensitivity model changes the $E_c$ limits by about 3 keV.}
\label{fig:20110213_chi2_hsi2}
\end{figure}

\begin{figure}[tbh]
\epsscale{1.0}
\plotone{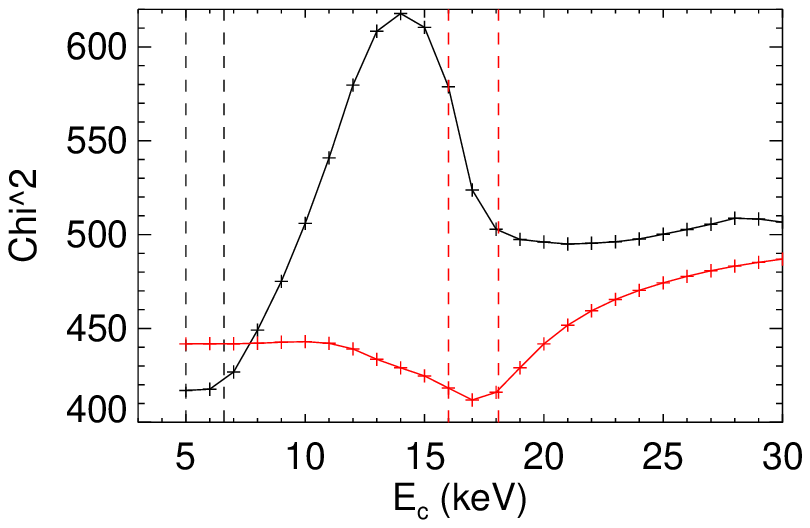}
\caption{$\chi^2$ vs. low cutoff energy $E_c$ for the time interval
  in Figure~\ref{fig:early_spec}, for coronal (black) and photospheric (red) 
  abundances. Vertical dashed lines denote upper and
  lower limits for $E_c$, defined as the points on the curve where
  $\chi^2(E_c)$ passes through $min(\chi^2)+6.63$, corresponding to the
  99\% confidence limit for the $\chi^2$ distribution.}
\label{fig:20110213_chi2_again}
\end{figure}

To attempt to include this sensitivity loss in the combined {\eve}+{\rhessi} calculations, we multiply the nominal {\rhessi} detector response by a factor of 0.5, reducing 
{\rhessi} sensitivity by a factor of 2 for the 2011
February 13 and 15 time intervals. We find that while the values of $\chi^2$ are not as good for the 
reduced-sensitivity solutions, there is not much change in the fit values of cutoff energy $E_c$. In Figure~\ref{fig:20110213_chi2_hsi2} we show the $\chi^2$ curves for the time interval 
2011 February 13, 17:32:36~UT for the reduced-sensitivity case. In Figure~\ref{fig:20110213_chi2_again} we reproduce Figure~\ref{fig:20110213_chi2} for comparison. As 
can be seen, the values for the limits 
on the cutoff energy $E_c$ do not change by much. This is a much smaller effect than the effect of changing
abundances as shown in the paper; hence we did not test or adopt the reduced-sensitivity model for the full sample of
flares. It is possible that a change in sensitivity that is energy-dependent may change the $E_c$
result, but we currently have no good way to test this.

\newpage
\bibliography{ppr2017}
\bibliographystyle{aasjournal.bst}
\newpage

\begin{table}[!htb]
\resizebox{\textwidth}{!}{%
\begin{tabular}{lcccccccccc}\hline
Date\_time & $\delta t$& $E_c$(coronal) & $E_c$(photo) & Log(\etotal) coronal & Log(\etotal) photo & Log(\ediff) & $\chi^2$ coronal & $\chi^2$ photo & $R(Fe)$ & $R_f(Fe)$\\
\hline
20110213\_173136& 60.0&$ 5.0 - 10.5$&$ 5.0 - 18.4$&$ 28.4 -  29.6$&$ 27.4 -  28.4$&$ 1.2$&$ 0.32$&$ 0.32$&$ 0.62$&$ 17.20$ \\
20110213\_173236& 60.0&$ 5.0 -  6.6$&$16.0 - 18.1$&$ 30.4 -  31.1$&$ 28.3 -  28.5$&$ 2.6$&$ 0.68$&$ 0.67$&$ 0.61$&$ 19.41$ \\
\\
20110215\_014712& 60.0&$ 5.0 - 11.2$&$ 5.0 - 30.0$&$ 28.4 -  30.3$&$ 20.9 -  21.4$&$ 2.5$&$ 0.52$&$ 0.51$&$ 0.50$&$ 15.47$ \\
20110215\_014812& 60.0&$ 5.0 -  7.3$&$15.2 - 18.4$&$ 30.4 -  31.2$&$ 28.5 -  28.7$&$ 1.7$&$ 0.20$&$ 0.18$&$ 0.51$&$ 13.94$ \\
\\
20110216\_142304& 60.0&$ 5.0 - 30.0$&$15.4 - 18.8$&$ 26.9 -  27.4$&$ 27.9 -  28.1$&$-0.0$&$ 0.15$&$ 0.22$&$ 0.96$&$ 24.67$ \\
20110216\_142404& 36.0&$ 5.0 -  9.3$&$16.1 - 17.7$&$ 29.3 -  31.1$&$ 28.1 -  28.3$&$ 2.9$&$ 0.15$&$ 0.20$&$ 0.79$&$ 30.61$ \\
\\
20110228\_124644& 60.0&$ 5.0 - 11.2$&$ 5.0 - 30.0$&$ 28.6 -  29.8$&$ 26.5 -  28.2$&$ 1.5$&$ 0.58$&$ 0.56$&$ 0.47$&$ 14.06$ \\
20110228\_124744& 60.0&$ 6.7 - 10.3$&$15.8 - 20.0$&$ 29.1 -  30.1$&$ 27.8 -  27.9$&$ 1.3$&$ 0.50$&$ 0.51$&$ 0.61$&$ 16.91$ \\
\\
20110307\_194720& 60.0&$ 5.0 - 12.1$&$ 5.0 - 30.0$&$ 28.2 -  29.8$&$ 25.6 -  25.7$&$ 2.0$&$ 0.20$&$ 0.19$&$ 0.58$&$ 21.81$ \\
20110307\_194820& 28.0&$ 5.0 - 12.2$&$ 5.0 - 30.0$&$ 28.3 -  29.4$&$ 25.7 -  28.4$&$ 0.8$&$ 0.22$&$ 0.22$&$ 0.67$&$ 21.12$ \\
\\
20110307\_214644& 60.0&$ 5.0 -  7.4$&$ 5.0 - 30.0$&$ 29.6 -  30.4$&$ 26.7 -  27.6$&$ 2.1$&$ 0.92$&$ 0.86$&$ 0.54$&$ 16.54$ \\
20110307\_214744& 60.0&$ 6.2 -  8.5$&$17.7 - 19.6$&$ 29.7 -  30.6$&$ 28.1 -  28.3$&$ 1.9$&$ 0.42$&$ 0.41$&$ 0.63$&$ 19.32$ \\
\\
20110308\_022736& 60.0&$ 5.0 -  6.8$&$ 5.0 - 21.2$&$ 30.2 -  31.0$&$ 27.8 -  30.9$&$ 2.8$&$ 0.35$&$ 0.39$&$ 0.59$&$ 21.72$ \\
20110308\_181044& 60.0&$ 5.0 - 30.0$&$ 5.0 - 24.0$&$ 27.3 -  28.4$&$ 27.7 -  28.4$&$ 0.4$&$ 0.47$&$ 0.49$&$ 0.49$&$ 13.00$ \\
\\
20110309\_135740& 60.0&$ 5.0 - 30.0$&$16.2 - 19.1$&$ 26.1 -  29.6$&$ 27.6 -  27.8$&$ 1.1$&$ 0.43$&$ 0.57$&$ 0.94$&$ 30.72$ \\
\\
20110309\_231820& 60.0&$ 5.0 -  9.0$&$ 5.0 - 30.0$&$ 29.6 -  31.1$&$ 25.3 -  25.5$&$ 1.6$&$ 0.29$&$ 0.36$&$ 0.67$&$ 22.24$ \\
20110309\_231920& 32.0&$22.0 - 26.5$&$15.6 - 21.1$&$ 28.0 -  28.5$&$ 28.6 -  28.8$&$-1.0$&$ 0.47$&$ 0.49$&$ 0.63$&$ 19.70$ \\
\\
20110314\_194952& 60.0&$ 5.0 -  8.1$&$16.7 - 18.2$&$ 29.8 -  31.1$&$ 28.4 -  28.5$&$ 1.8$&$ 0.44$&$ 0.52$&$ 0.65$&$ 21.09$ \\
20110314\_195108& 60.0&$ 5.0 - 29.0$&$ 5.0 - 25.9$&$ 27.9 -  28.6$&$ 28.0 -  28.9$&$-1.3$&$ 0.36$&$ 0.45$&$ 0.58$&$ 23.63$ \\
\\
20110315\_002104& 60.0&$ 5.0 -  7.6$&$17.7 - 20.2$&$ 29.6 -  30.7$&$ 27.8 -  27.9$&$ 2.7$&$ 1.10$&$ 1.17$&$ 0.71$&$ 24.20$ \\
\\
20110422\_043824& 60.0&$19.6 - 28.6$&$13.9 - 23.2$&$ 26.6 -  27.1$&$ 27.0 -  27.5$&$-0.9$&$ 0.74$&$ 0.76$&$ 0.94$&$ 29.25$ \\
20110422\_043924& 28.0&$ 5.0 - 30.0$&$14.7 - 26.3$&$ 26.1 -  30.0$&$ 26.6 -  27.4$&$ 2.4$&$ 0.41$&$ 0.51$&$ 1.13$&$ 40.01$ \\
\\
20110529\_101236& 56.0&$ 5.0 - 30.0$&$17.3 - 20.0$&$ 16.9 -  28.4$&$ 27.5 -  27.7$&$ 0.5$&$ 0.37$&$ 0.49$&$ 1.01$&$ 37.38$ \\
\\
20110607\_062044& 60.0&$ 8.7 - 10.3$&$20.9 - 30.0$&$ 27.8 -  28.0$&$ 26.5 -  26.5$&$ 0.9$&$ 0.63$&$ 0.96$&  NA &$ 20.11$ \\
20110607\_062144& 48.0&$ 8.2 -  9.2$&$25.8 - 28.2$&$ 28.3 -  28.5$&$ 26.8 -  26.8$&$ 1.1$&$ 0.70$&$ 1.03$&  NA &$ 19.84$ \\
\\
20110730\_020652& 36.0&$ 5.0 - 11.6$&$17.4 - 27.7$&$ 27.5 -  28.3$&$ 26.8 -  27.0$&$ 0.4$&$ 0.36$&$ 0.50$&  NA &$ 11.09$ \\
20110730\_020732& 24.0&$20.4 - 29.0$&$19.4 - 22.3$&$ 28.2 -  28.3$&$ 28.4 -  28.5$&$-0.7$&$ 1.00$&$ 1.32$&$ 0.84$&$ 30.54$ \\
\\
20110803\_033428& 52.0&$19.3 - 22.4$&$17.0 - 18.1$&$ 27.2 -  27.4$&$ 28.0 -  28.2$&$-0.9$&$ 0.74$&$ 1.28$&$ 0.86$&$ 34.57$ \\
\\
20110803\_043016& 44.0&$ 5.0 - 13.6$&$18.0 - 27.0$&$ 27.2 -  27.8$&$ 26.9 -  27.0$&$ 0.3$&$ 0.27$&$ 0.50$&  NA &$  9.28$ \\
\\
20110809\_080028& 60.0&$15.0 - 25.6$&$17.8 - 21.7$&$ 27.2 -  27.8$&$ 27.6 -  27.8$&$-0.2$&$ 0.41$&$ 1.01$&$ 1.10$&$ 42.21$ \\
\end{tabular}}
\end{table}

\begin{table}[!htb]
\resizebox{\textwidth}{!}{%
\begin{tabular}{lcccccccccc}\hline
Date\_time & $\delta t$&$E_c$(coronal) & $E_c$(photo) & Log(\etotal) coronal & Log(\etotal) photo & Log(\ediff) & $\chi^2$ coronal & $\chi^2$ photo & $R(Fe)$ & $R_f(Fe)$\\
\hline
20110906\_013736& 60.0&$ 5.0 - 21.4$&$ 5.0 - 21.1$&$ 27.4 -  30.8$&$ 27.5 -  30.9$&$ 2.9$&$ 1.53$&$ 1.80$&$ 0.62$&$ 22.44$ \\
20110906\_013836& 40.0&$ 5.0 -  6.4$&$ 5.0 -  6.1$&$ 30.1 -  30.8$&$ 30.6 -  31.1$&$ 2.8$&$ 1.32$&$ 1.60$&$ 0.73$&$ 30.70$ \\
\\
20110906\_221556& 60.0&$20.9 - 22.6$&$19.7 - 20.3$&$ 27.5 -  27.6$&$ 27.8 -  27.8$&$-1.0$&$ 0.88$&$ 1.09$&$ 0.87$&$ 37.28$ \\
\\
20110908\_153616& 36.0&$ 6.2 - 10.4$&$ 5.0 - 30.0$&$ 27.9 -  28.8$&$ 25.8 -  25.9$&$ 1.2$&$ 0.95$&$ 1.13$&  NA &$ 30.86$ \\
20110908\_153656& 60.0&$ 5.0 - 30.0$&$ 5.0 - 30.0$&$ 26.6 -  26.8$&$ 26.6 -  26.7$&$-0.7$&$ 0.51$&$ 0.65$&$ 1.37$&$ 63.40$ \\
\\
20110924\_171944& 60.0&$21.3 - 25.3$&$18.3 - 22.4$&$ 27.6 -  27.7$&$ 27.6 -  28.0$&$-0.8$&$ 1.11$&$ 1.09$&$ 0.71$&$ 30.35$ \\
20110924\_172044& 60.0&$ 5.0 -  5.7$&$ 5.0 -  5.3$&$ 31.6 -  31.9$&$ 31.8 -  31.9$&$ 3.1$&$ 1.10$&$ 1.44$&$ 0.60$&$ 26.81$ \\
\\
20110924\_191052& 60.0&$19.0 - 24.4$&$21.7 - 24.5$&$ 27.4 -  27.6$&$ 27.5 -  27.6$&$-0.5$&$ 0.96$&$ 1.16$&$ 0.97$&$ 49.26$ \\
\\
20110924\_203448& 60.0&$10.8 - 11.4$&$21.9 - 28.4$&$ 30.5 -  30.6$&$ 28.0 -  28.2$&$ 0.8$&$ 0.57$&$ 0.45$&$ 0.41$&$ 17.49$ \\
\\
20110926\_050544& 40.0&$ 5.0 -  5.5$&$ 5.0 -  6.1$&$ 29.5 -  29.8$&$ 29.1 -  29.6$&$ 2.3$&$ 2.26$&$ 2.20$&  NA &$  5.67$ \\
\\
20111002\_004144& 60.0&$16.1 - 24.8$&$18.5 - 23.9$&$ 26.9 -  27.3$&$ 27.1 -  27.2$&$-0.3$&$ 0.91$&$ 1.15$&$ 1.07$&$ 54.35$ \\
\\
20111105\_030920& 60.0&$17.7 - 20.0$&$17.9 - 19.3$&$ 27.4 -  27.7$&$ 27.5 -  27.7$&$-0.6$&$ 0.41$&$ 0.93$&$ 1.24$&$ 47.97$ \\
\\
20111226\_021836& 60.0&$ 5.0 - 30.0$&$ 5.0 - 30.0$&$ 25.8 -  26.0$&$ 26.1 -  26.7$&$-0.4$&$ 0.36$&$ 0.53$&$ 1.45$&$ 70.60$ \\
\\
20111226\_201544& 60.0&$18.7 - 27.8$&$18.8 - 24.3$&$ 27.1 -  27.3$&$ 27.4 -  27.5$&$-0.2$&$ 1.19$&$ 1.50$&$ 1.05$&$ 41.18$ \\
\\
20111231\_161908& 60.0&$ 5.0 -  7.0$&$16.8 - 23.0$&$ 28.4 -  28.8$&$ 26.8 -  27.0$&$ 1.6$&$ 1.65$&$ 1.88$&  NA &$ 13.39$ \\
20111231\_162008& 28.0&$ 8.0 - 10.2$&$ 5.0 - 30.0$&$ 28.0 -  28.4$&$ 25.7 -  25.9$&$ 1.2$&$ 1.32$&$ 1.71$&  NA &$ 25.74$ \\
\\
20121113\_054512& 36.0&$ 5.8 -  7.0$&$ 6.2 -  7.4$&$ 29.2 -  29.5$&$ 28.6 -  29.0$&$ 2.2$&$ 0.49$&$ 0.47$&  NA &$  9.34$ \\
20121113\_054556& 60.0&$ 5.0 -  6.1$&$ 5.0 -  6.1$&$ 30.6 -  31.1$&$ 30.5 -  31.0$&$ 2.9$&$ 0.56$&$ 0.63$&$ 0.70$&$ 29.89$ \\
\\
20121114\_040052& 60.0&$ 5.0 - 11.0$&$ 5.0 - 30.0$&$ 27.2 -  28.1$&$ 26.3 -  27.2$&$ 1.0$&$ 0.10$&$ 0.10$&  NA &$  5.23$ \\
20121114\_040152& 40.0&$ 5.9 -  6.1$&$16.5 - 22.5$&$ 29.2 -  29.2$&$ 27.0 -  27.1$&$ 2.1$&$ 1.20$&$ 1.21$&  NA &$ 12.38$ \\
\\
20121120\_123808& 32.0&$ 5.7 -  6.0$&$ 6.1 -  8.8$&$ 29.6 -  29.7$&$ 28.1 -  29.0$&$ 2.3$&$ 1.03$&$ 0.82$&  NA &$ 27.42$ \\
20121120\_123844& 60.0&$ 5.0 -  6.0$&$ 5.0 -  6.1$&$ 30.8 -  31.2$&$ 30.4 -  30.9$&$ 2.8$&$ 0.54$&$ 0.67$&$ 0.74$&$ 28.72$ \\
\\
20121121\_064812& 44.0&$ 6.9 -  7.0$&$ 7.8 - 26.8$&$ 29.3 -  29.3$&$ 27.1 -  28.5$&$ 1.8$&$ 0.79$&$ 0.78$&  NA &$ 10.81$ \\
20121121\_064900& 60.0&$ 5.0 -  8.2$&$16.5 - 20.1$&$ 29.5 -  30.7$&$ 27.6 -  28.0$&$ 1.8$&$ 0.99$&$ 0.96$&$ 0.57$&$ 19.97$ \\
\\
20121128\_213200& 60.0&$ 5.0 -  6.3$&$17.0 - 18.2$&$ 30.8 -  31.4$&$ 28.2 -  28.3$&$ 2.9$&$ 0.32$&$ 0.51$&$ 0.72$&$ 29.01$ \\
20121128\_213300& 60.0&$ 5.0 -  6.5$&$ 5.0 - 30.0$&$ 30.9 -  31.6$&$ 26.0 -  26.2$&$ 2.9$&$ 0.39$&$ 0.99$&$ 0.76$&$ 29.40$ \\
\\
20130111\_085804& 52.0&$ 5.0 -  7.0$&$19.4 - 24.0$&$ 28.5 -  29.0$&$ 26.9 -  27.2$&$ 1.3$&$ 2.15$&$ 2.76$&  NA &$ 14.26$ \\
20130111\_085900& 60.0&$ 5.0 - 30.0$&$13.9 - 17.5$&$ 25.3 -  28.2$&$ 27.6 -  27.8$&$ 0.4$&$ 2.41$&$ 2.36$&$ 0.75$&$ 25.57$ \\
\\
20130113\_004744& 48.0&$ 5.0 - 14.1$&$14.7 - 30.0$&$ 27.1 -  27.5$&$ 26.8 -  26.9$&$ 0.2$&$ 0.30$&$ 0.41$&  NA &$  5.21$ \\
20130113\_004840& 60.0&$ 5.0 -  9.3$&$16.6 - 17.8$&$ 29.1 -  30.9$&$ 28.2 -  28.3$&$ 2.0$&$ 0.62$&$ 0.89$&$ 0.78$&$ 26.82$ \\
\\
20130217\_154704& 60.0&$ 5.0 -  7.1$&$16.5 - 19.1$&$ 29.7 -  30.7$&$ 27.5 -  27.7$&$ 2.8$&$ 0.70$&$ 0.94$&$ 0.88$&$ 33.21$ \\
\end{tabular}}
\caption{Fitting results. Values (columns) for each time interval (row) are:
interval date and time;
derived limits for $E_c$ using coronal and photospheric abundances;
derived limits for {\etotal} (total electron energy) for coronal and photospheric abundances;
minimum (best-fit) values of \textit{reduced} $\chi^2$ using coronal and photospheric abundances;
difference between {\etotal} values for the derived $E_c$ (coronal) and $E_c = 15$~keV;
$R(Fe)$, the ratio of the count rate in the $\sim$6.7~keV Fe line complex to the
peak of the count rate in the 10--20~keV range; and
$R_f(Fe)$, the ratio of the fitted photon flux in the Fe line complex to the photon flux 
in the 10--20~keV range. For $R(Fe)$, ``NA'' means that there was no 
separate peak in the count spectrum, due to the fact that the thin 
attenuator was out.}

\label{table1}
\end{table}

\begin{table}[!htb]
\resizebox{\textwidth}{!}{%
\begin{tabular}{lcccccc}\hline
Date\_time & $Log(EM_{16})$ (cm$^{-3}$) & $Log(N_{16})$ (cm$^{-3}$) & 
$Log(V_{16})$ (cm$^3$) & $\tau_{Fe\textsc{xxv}}$ (s) & 
$\tau_\textsc{heat}$ (s) & $R_f(Fe, 2)/R_f(Fe,1)$\\
\hline
20110213\_173136,173236& 47.4& 10.4& 26.5&   5.&  49.&1.13 \\
20110215\_014712,014812& 47.4& 10.3& 26.8&   7.&  44.&0.90 \\
20110216\_142304,142404& 47.7& 10.5& 26.7&   4.&  71.&1.24 \\
20110228\_124644,124744& 47.4& 10.3& 26.8&   6.&  64.&1.20 \\
20110307\_194720,194820& 47.0&  9.9& 27.3&  18.& 137.&0.97 \\
20110307\_214644,214744& 47.3& 10.4& 26.6&   6.&  41.&1.17 \\
20110309\_231820,231920& 48.1& 10.4& 27.3&   5.&  70.&0.89 \\
20110314\_194952,195108& 48.6& 11.0& 26.6&   1.&  37.&1.12 \\
20110422\_043824,043924& 47.5& 10.5& 26.5&   4.&  83.&1.37 \\
20110607\_062044,062144& 45.9&  9.3& 27.3&  64.& 171.&0.99 \\
20110730\_020652,020732& 47.9& 10.5& 26.9&   4.&  30.&2.75 \\
20110906\_013736,013836& 47.4&  9.9& 27.5&  15.& 106.&1.37 \\
20110908\_153616,153656& 47.2& 10.2& 26.7&   8.&  31.&2.05 \\
20110924\_171944,172044& 47.6& 10.5& 26.5&   4.&  94.&0.88 \\
20111231\_161908,162008& 46.1&  9.8& 26.5&  20.& 180.&1.92 \\
20121113\_054512,054556& 47.0&  9.9& 27.2&  17.&  33.&3.20 \\
20121114\_040052,040152& 45.5&  9.2& 27.1&  79.&  32.&2.37 \\
20121120\_123808,123844& 47.3& 10.1& 27.2&  11.&  30.&1.05 \\
20121121\_064812,064900& 46.9&  9.8& 27.3&  21.&  37.&1.85 \\
20121128\_213200,213300& 48.1& 10.9& 26.4&   2.& 105.&1.01 \\
20130111\_085804,085900& 46.9&  9.9& 27.1&  16.&  44.&1.79 \\
20130113\_004744,004840& 47.6& 10.2& 27.1&   8.&  30.&5.15 \\
\end{tabular}}
\caption{Results for evaluating ionization equilibrium for flares where two consecutive time 
intervals could be fit, using volume estimates from {\rhessi} images. For each flare (row), 
values (columns) are:
$EM_{16}$, the total emission measure for plasma with $T>16$~MK ($Log(T)>7.2$), integrated from 
the DEM; $N_{16}$, the plasma density;
$V_{16}$, the volume estimated from {\rhessi} images;
$\tau_{Fe\textsc{xxv}}$, the time scale for ionization equilibrium of Fe~\textsc{xxv};  
$\tau_\textsc{heat}$, the heating time scale, and the ratio of $R_f(Fe)$, 
Fe line to continuum flux ratios, for the second interval to the first interval.}
\label{table2}
\end{table}

\end{document}